\documentclass[a4paper,12pt]{article}
\usepackage{amssymb}
\usepackage{latexsym}

\usepackage{url}
\usepackage{xcolor}
\usepackage{bm}
\definecolor{newcolor}{rgb}{.8,.349,.1}
\usepackage{graphicx}

\usepackage[thinlines,thiklines]{easybmat}

\begin{document}
\title{Solving the Faddeev-Merkuriev equations in total 
	orbital momentum representation via 
	spline collocation and tensor product preconditioning}
\author{Vitaly A. Gradusov \and Vladimir A. Roudnev \and Evgeny A. Yarevsky \and Sergey L. Yakovlev\thanks{Department of Computational Physics, St Petersburg State University}}
\date{}

\maketitle

\begin{abstract}
The computational
approach for solving the Faddeev-Merkuriev equations in total orbital 
momentum representation is presented. These equations describe a system of three quantum charged particles and are widely used in bound state and scattering calculations. The 
approach is based on the spline collocation method and exploits intensively 
the tensor product form of discretized operators and preconditioner, which leads to a 
drastic economy in both computer resources
and time.
\end{abstract}

%

\section{Introduction}

Since the pioneering work of Hylleraas~\cite{Hylleraas}, quantum three-body systems remain the source of challenges and inspirations for theoretical and experimental physicists. 
New effects specific to three-body systems have been predicted, such as Thomas effect~\cite{Thomas}, Efimov effect~\cite{Efimov}, the Phillips line~\cite{Phillips,Roudn12-2}. 
Direct modeling of nuclear and molecular three-body systems paved a way to develop and to fine-tune realistic models of inter-atomic and inter-nucleon 
interactions~\cite{Aziz91,Cybul99,Jezior07,Witala03}. 
Ab-initio calculations of some specific three-atomic systems may give essential contributions to metrology~\cite{Roudn12}. 
The Coulomb quantum three-body systems are also of great importance. For instance, delicate calculations of asymmetric heavy-hydrogen molecular ions  gave an insight on 
$mu$-catalysis~\cite{Belyaev96}, studies of positron-atom interactions are valuable for positron-emission tomography.

Even though the basic mathematical model for such a broad spectrum of physical systems is the Schr\"odinger equation, the diversity of 
model interactions and particular physical states 
leads to a variety of  employed computational methods~\cite{Bhatia74,Humber97,Mitroy95,Igara94, Ward12,Kvits95, Hu99, Papp-Yak01, Yak07, Lazau17,Volk09,Volk15,Yar15,Yar17}.
Thus, our ability to perform direct model-free calculations for such wide range of systems is of utmost importance for many branches 
of physics. 

Our goal is to present a universal and efficient computational framework applicable to this broad variety of physical systems and states. 
In order to achieve this goal we start from the following presuppositions. The approach should be based on a physically correct and mathematically sound representation of the problem.
The  Faddeev equations formalism~\cite{Fadd93} fulfills all of these requirements.
Clear separation of asymptotic channels corresponding to different clusterisations of the system 
is one of the main advantageous features of the formalism from the point of view of practical applications.
Coulomb systems are incorporated into the original  formalism by the Merkuriev's version of the Faddeev equations~\cite{Merkur80, Fadd93}.
Being mathematically equivalent to the Schr\"odinger equations, the Faddeev-Merkuriev (FM) equations have advantages of much simpler boundary conditions and much simpler behavior of their solutions. 
This leads to much weaker requirements for the basis employed in the calculations.

Direct solution of the FM
equations is not, however, a simple task. In order to reduce the dimensionality of the configuration space the symmetries of the solutions must be taken into account.
We base our computational approach on total orbital momentum representation which leads to systems of partial differential equations in three-dimensional space. Solving such systems numerically 
is still a challenging task which calls for developing an effective and robust preconditioning technique. Here we propose a preconditioning scheme based on the tensor-trick algorithm 
and compare our numerical scheme with solving the corresponding sparse linear system using PARDISO direct solver. 
Our approach clearly outperforms the direct method both in time and memory requirements, which paves a way to accurate calculations of rather challenging systems, 
including highly rotationally excited three-body states.

In the following sections we give a description of the FM equations formalism, describe our numerical scheme, and give a few computational examples 
for some well-studied systems of diverse physical nature. 

Throughout the paper we assume $\hbar =1$ and we use bold font for vectors as, for instance, $\bm{x}$ and normal font for their magnitude $x=|\bm{x}|$.
 
\section{The Faddeev-Merkuriev equations}
\label{FMeq}

\subsection{Notation and basic equations}

\begin{figure}[!t]
\centering
\includegraphics[scale=.5]{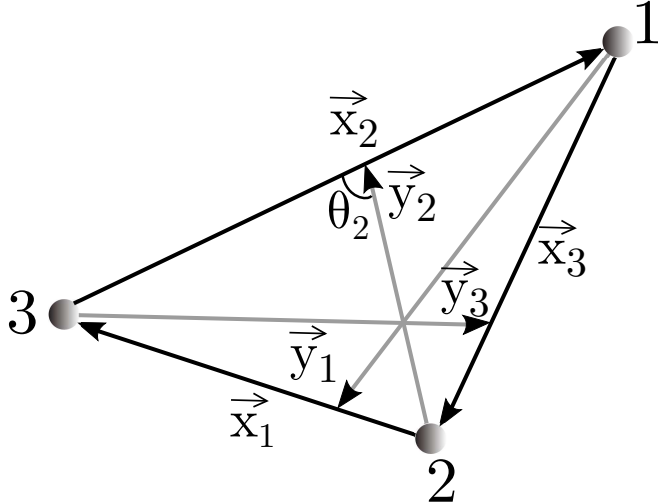}
\caption{Jacobi coordinates for three particles.}
\label{Jacobi}
\end{figure}

The FM equations 
for three quantum particles
are
of the form
\begin{eqnarray}
\label{FM-6d}
\{ T_\alpha  +  V_\alpha(x_\alpha) +
\sum_{\beta\ne\alpha}V_\beta^{(\mathrm{l})}(x_\beta, y_\beta) &-& E \} \psi_\alpha(\bm{x_\alpha}, \bm{y_\alpha}) = \nonumber \\
  &-&  V_\alpha^{(\mathrm{s})}(x_\alpha,y_\alpha) \sum_{\beta\ne\alpha}\psi_\beta(\bm{x_\beta}, \bm{y_\beta}),\quad\alpha=1,2,3.
\end{eqnarray}
These equations describe bound and scattering states with
the  
energy $E$ of a system of three spinless non-relativistic charged particles of masses $m_\alpha$ and charges $Z_\alpha$ in their center of mass system of coordinates~\cite{Fadd93}.
In what follows the indices $\alpha$, $\beta$, $\gamma$ run over the set $\{1,2,3\}$ enumerating particles and an index, say $\alpha$,
is also used for identifying the complementary pair of particles,
since
in the partition $\{\alpha(\beta \gamma)\}$ the pair of particles $\beta\gamma$ is uniquely determined by the particle with the number $\alpha$.
The particle positions are described by pairs of Jacobi vectors (see Fig.~\ref{Jacobi}).
The standard Jacobi coordinates are defined for a partition $\alpha(\beta\gamma)$ as the relative position vector $\bm{x}_\alpha$ between the particles of the pair $\beta\gamma$ and the relative vector $\bm{y}_\alpha $ between their center of mass and the particle $\alpha$. In applications it is convenient to use reduced Jacobi coordinates $\bm{x}_\alpha, \bm{y}_\alpha$ which are Jacobi vectors scaled by factors $\sqrt{2\mu_{\alpha}}$ and $\sqrt{2\mu_{\alpha(\beta\gamma)}}$, respectively, where the reduced masses are given by
\begin{equation}
\mu_{\alpha} = \frac{m_\beta m_\gamma}{m_\beta+m_\gamma},\qquad \mu_{\alpha(\beta\gamma)}=\frac{m_\alpha (m_\beta+m_\gamma)}{m_\alpha+m_\beta+m_\gamma}.
\end{equation}
For different $\alpha's$ the reduced Jacobi vectors are related by an orthogonal transform
\begin{equation}
\bm{x}_\beta=c_{\beta\alpha}\bm{x}_\alpha + s_{\beta\alpha}\bm{y}_\alpha, \qquad
\bm{y}_\beta=-s_{\beta\alpha}\bm{x}_\alpha + c_{\beta\alpha}\bm{y}_\alpha,
\label{JacobiTrans}
\end{equation}
where
$$
c_{\beta\alpha}=-\left[\frac{m_\beta m_\alpha}{(M-m_\beta)(M-m_\alpha)}\right]^{1/2}, \qquad
s_{\beta\alpha}=(-1)^{\beta-\alpha}\mbox{sgn}(\alpha-\beta)(1-c^2_{\beta\alpha})^{1/2}
$$
and $M=\sum_\alpha m_\alpha$.
In what follows, where it is due, it is assumed that $\beta$ Jacobi vectors are represented through $\alpha$ vectors via (\ref{JacobiTrans}).
The orientation of Jacobi vectors is important, and it is chosen as it is shown in Fig.~\ref{Jacobi}.
The kinetic energy operators are given by $T_\alpha\equiv-\Delta_{\bm{x_\alpha}} - \Delta_{\bm{y_\alpha}}$.
The potentials $V_\alpha$ represent a sum of the pairwise Coulomb interaction $V_{\alpha}^{\mathrm{C}}(x_\alpha)=\sqrt{2\mu_{\alpha}}Z_\beta Z_\gamma/x_\alpha$ ($\beta,\gamma\ne\alpha$) and a short-range $V_{\alpha}^{\mathrm{sh}}(x_\alpha)$ (decreasing faster than $1/x^2_{\alpha}$ as $x_\alpha \to \infty$) potential.
The potentials $V_{\alpha}$ are split into the interior (short-range) $V^{(\mathrm{s})}_\alpha$ and the tail (long-range) parts $V^{(\mathrm{l})}_\alpha$
\begin{equation}
\label{PotSplit}
V_{\alpha}(x_\alpha) = V^{(\mathrm{s})}_\alpha(x_\alpha,y_\alpha) + V^{(\mathrm{l})}_\alpha(x_\alpha, y_\alpha).
\end{equation}
The equations~(\ref{FM-6d}) can be summed up leading to the Schr\"odinger equation for the wave function
$\Psi=\sum_{\alpha}\psi_\alpha$, where $\psi_{\alpha}$ are the wave function components given by the solution of equations~(\ref{FM-6d}).

The splitting procedure~(\ref{PotSplit}) is especially important for scattering calculations. It makes the properties of the FM equations for the Coulomb potentials as appropriate for scattering problems as the standard Faddeev equations in the case of short-range potentials \cite{Fadd93,Papp-Yak01}. The key property of the FM equations~(\ref{FM-6d})
is that the right-hand side of each equation is a square-integrable function (i.e. confined to the vicinity of the triple collision point) \cite{Fadd93,Yak-Papp10}.
Splitting~(\ref{PotSplit}) of the potentials in general case is done in the three-body configuration space by the Merkuriev cut-off function $\chi_\alpha$ \cite{Fadd93}
\begin{equation}
V_\alpha^{(\mathrm{s})}(x_\alpha,y_\alpha) = \chi_\alpha(x_\alpha, y_\alpha) V_\alpha^{\mathrm{C}}(x_\alpha)+V_\alpha^{\mathrm{sh}}(x_\alpha),\quad V_\alpha^{(\mathrm{l})}(x_\alpha,y_\alpha) = \left(1-\chi_\alpha(x_\alpha, y_\alpha)\right) V_\alpha^{\mathrm{C}}(x_\alpha).
\label{split3b}
\end{equation}
 This splitting  confines  the short-range part of the potential to the regions in the three-body configuration space corresponding to the three-body collision point (particles are close to each other) and the binary configuration ($x_\alpha \ll y_\alpha$, as $y_\alpha \to \infty$). The form of the cut-off function can be rather arbitrary within some general requirements~\cite{Merkur80}. 
 One of the most often used variant~\cite{Kvits92} of this function reads
\begin{equation}
\label{Mcutoff0}
\chi_\alpha(x_\alpha, y_\alpha) =
2/\left\{1+\exp[ (x_\alpha/x_{0\alpha})^{\nu_\alpha}/(1+y_\alpha/y_{0\alpha}) ]\right\}.
\end{equation}
The parameters $x_{0\alpha}$, $y_{0\alpha}$ and $\nu_\alpha>2$ can in principle be chosen arbitrarily, but their choice changes the properties of components $\psi_\alpha$ that are important from both 
theoretical and computational points of view \cite{Yak-Papp10}.
In the papers~\cite{Grad19, Grad16} we have discussed the choice of the cut-off function and have presented some practical algorithm of how to choose its parameters efficiently.
In the bound state calculations the splitting can be omitted, that is one can take $V_\alpha^{(\mathrm{s})} = V_\alpha$ and $V_\alpha^{(\mathrm{l})} = 0$. Then the equations~(\ref{FM-6d}) turn into standard Faddeev equations.
On the other hand, in many calculations the choice $V_\beta^{(\mathrm{s})}=0$, $V_\beta^{(\mathrm{l})}=V_\beta$ with $\beta=3$ or $\beta=2,3$ is possible~\cite{Grad19}.
Then the set of equations~(\ref{FM-6d}) decouples leading to two equations in the first case, and to the Schr\"odinger equation in the second case.
Therefore in principle, the methods of this paper are
applicable to solving the Schr\"odinger equation as well.

Asymptotic boundary conditions are to be added to equations~(\ref{FM-6d}).
The bound state calculations imply zero Dirichlet-type  
boundary conditions $\psi_\alpha(\bm{x_\alpha},\bm{y_\alpha})\to 0$ as $x_\alpha$ or $y_\alpha\to\infty$.
The resulting eigenvalue problem allows one to determine the discrete energy spectrum.
The scattering problem gives rise to much more complicated radiation boundary conditions and a subsequent boundary value problem.
The exact asymptotic
form
of each component depends on the total energy $E$.
Comprehensive descriptions of scattering boundary conditions  can be found in
\cite{Fadd93}
(for the total energy $E$ of the system below the three-body breakup threshold see also  formula~(8) of~\cite{Grad19} and discussion therein).
Summarizing, for equations~(\ref{FM-6d}) 
two problems are possible:
\begin{itemize}
\item determination of discrete energy levels of bound states: eigenvalue problem with equations~(\ref{FM-6d}) and asymptotic zero Dirichlet-type  boundary conditions
\item scattering problem at a given total energy $E$: boundary value problem with equations~(\ref{FM-6d}) and asymptotic boundary conditions for scattering states.  
\end{itemize}
The methods that we describe in this paper are equally applicable for solving both problems, but for definiteness in what follows we refer to the bound state problem.

\subsection{Total orbital momentum representation}

Equations~(\ref{FM-6d}) are six dimensional PDE and therefore their direct solving on the modern computers is quite a challenging problem.   
One of the ways  to decrease the dimension of the problem is to use some basis expansion for the solution $\psi_\alpha$. 
Among the approaches used for
calculations, an expansion in bipolar harmonics is often used~\cite{Schelling89, Hu99, Lazau03}, which reduces~(\ref{FM-6d}) to an infinite set of coupled two dimensional integro-differential equations.
This set is then truncated to the finite set of equations to make their numerical solution possible.
In~\cite{Kostr89} another approach was proposed.
It is based on the expansion of components $\psi_\alpha$ in terms of eigenfunctions of the total
orbital momentum operator.
The total
orbital momentum is an integral of motion for the three-particle system.  
This makes it possible to reduce the FM equations to a finite set of three dimensional (3D) PDE by projecting onto a subspace of a given total
orbital momentum.
These equations are the 3D FM equations in total 
orbital momentum representation and are the equations that we solve in this work.
In the subsequent
 part of this section we briefly outline the derivation of the 3D FM equations 
 by using a notation more appropriate for our purposes than in~\cite{Kostr89}.

\begin{figure}[!t]
\centering
\begin{minipage}[h]{0.49\textwidth}
\center{\includegraphics[width=1\textwidth]{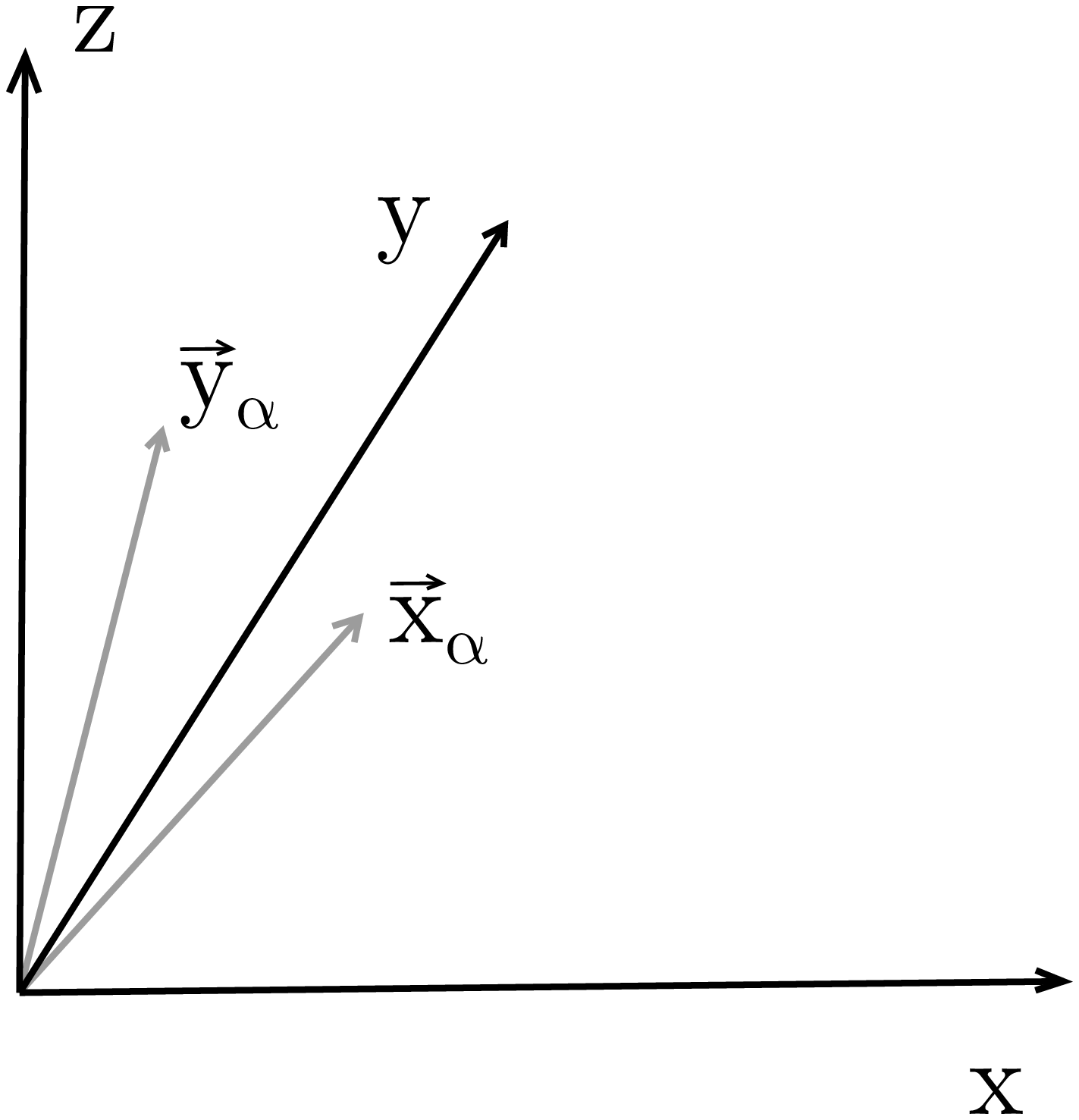}} \\ (a)
\end{minipage}
\hfill
\begin{minipage}[h]{0.49\textwidth}
\center{\includegraphics[width=1\textwidth]{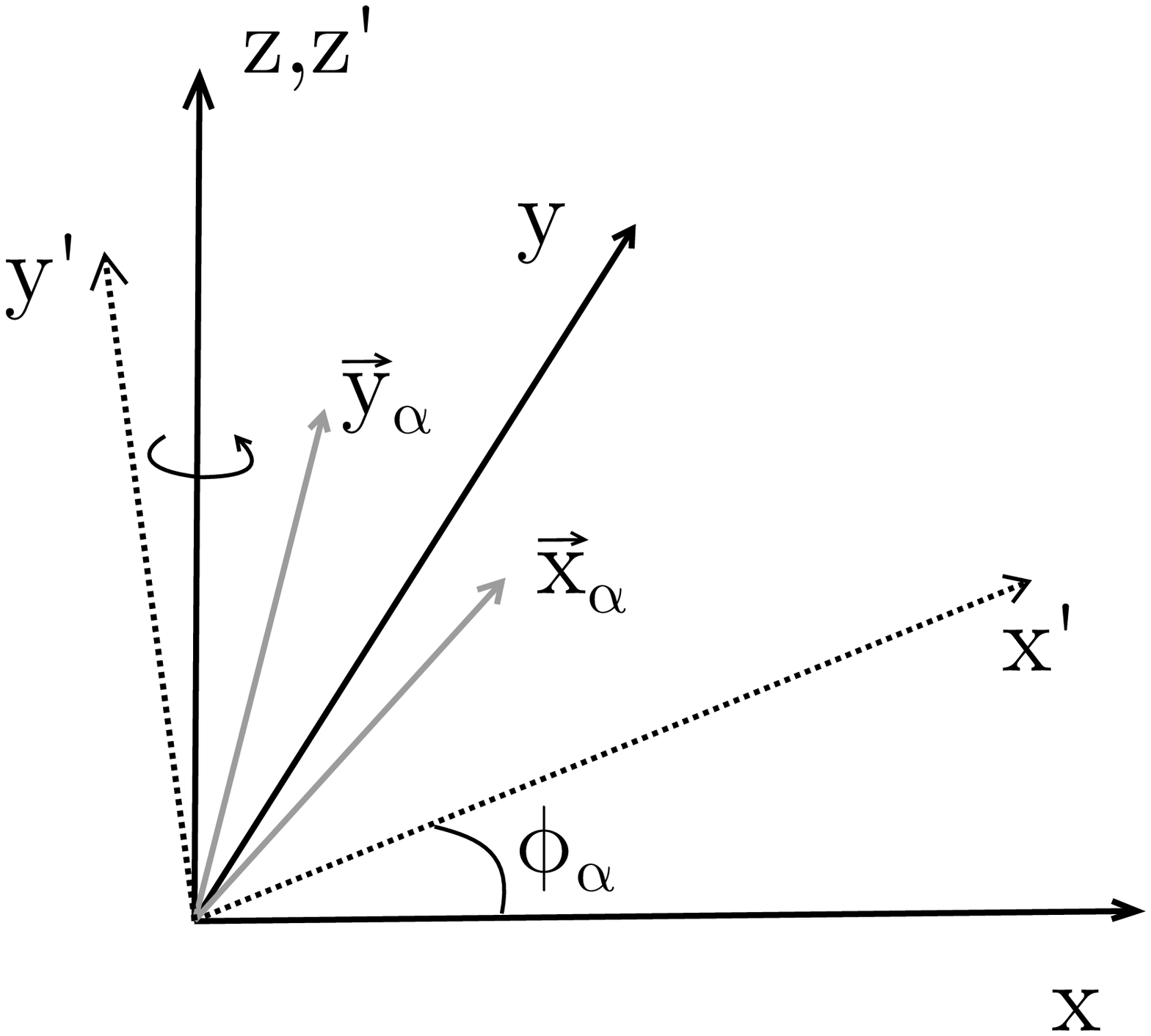}} \\ (b)
\end{minipage}
\hfill
\begin{minipage}[h]{0.49\textwidth}
\center{\includegraphics[width=1\textwidth]{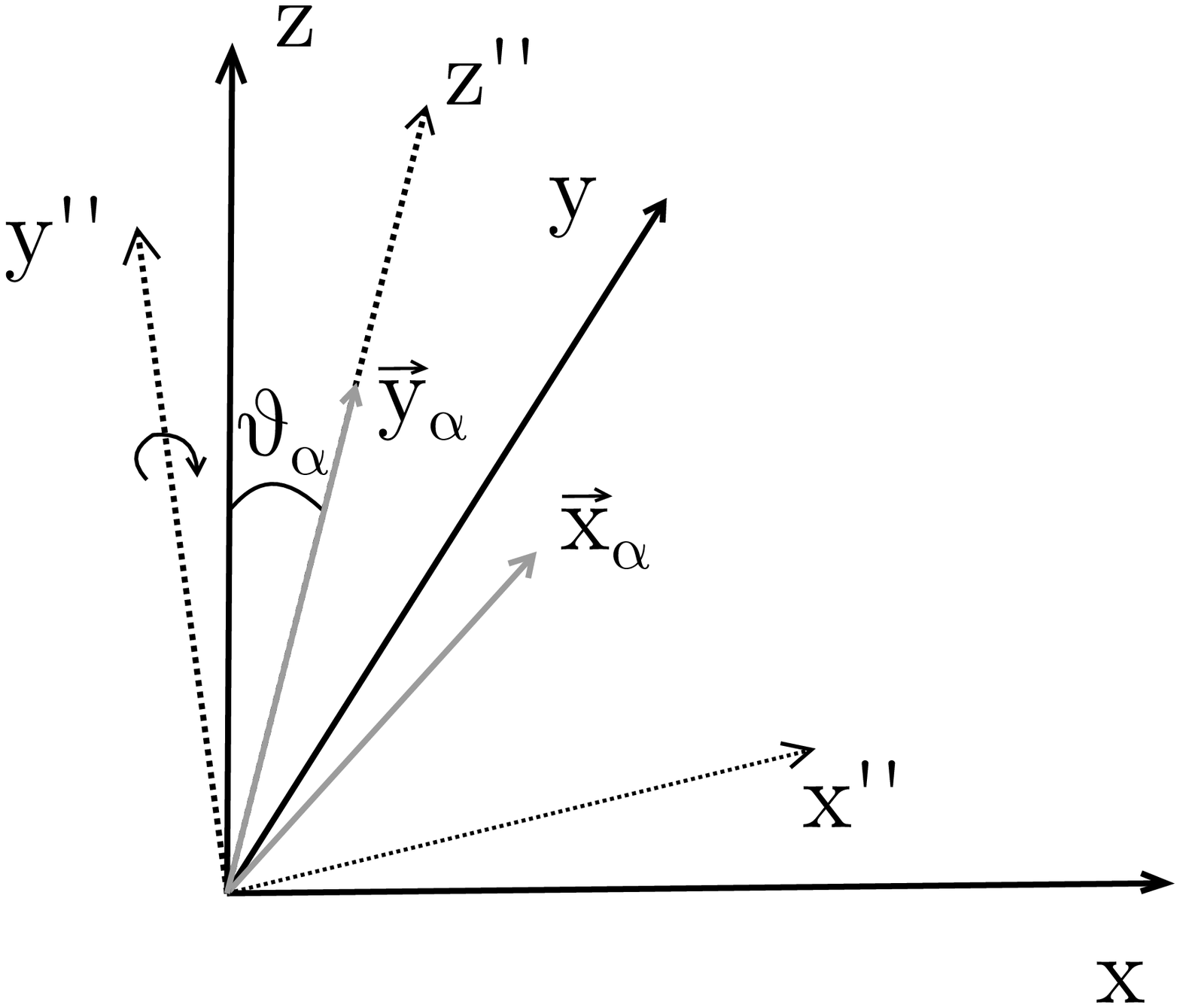}} \\ (c)
\end{minipage}
\hfill
\begin{minipage}[h]{0.49\textwidth}
\center{\includegraphics[width=1\textwidth]{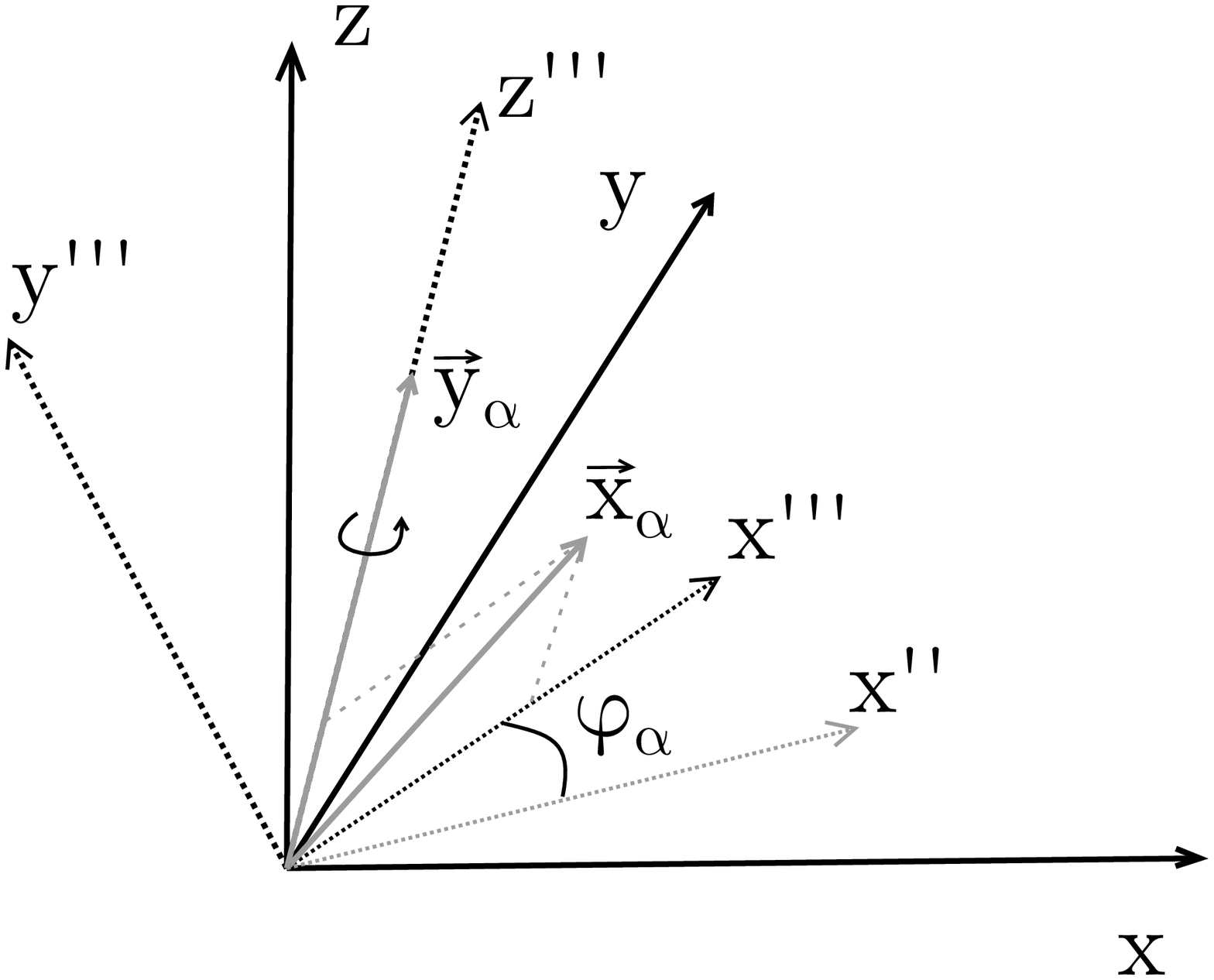}} \\ (d)
\end{minipage}
\caption{Definition of Euler angles $\phi_\alpha,\vartheta_\alpha,\varphi_\alpha$.}.
\label{Angles}
\end{figure}

We start by
introducing new kinematic coordinates ($X_\alpha$, $\Omega_\alpha$) in the six dimensional configuration space of the problem.
The coordinates $X_\alpha=\{x_\alpha,y_\alpha,z_\alpha\}$ determine particle positions in the plane which contains all three particles,
 $z_\alpha\equiv
 (\bm{x_\alpha},\bm{y_\alpha})/(x_\alpha y_\alpha)$ is cosine of an angle $\theta_\alpha$ between the vectors $\bm{x_\alpha}$ and $\bm{y_\alpha}$.
They are depicted in Fig.~\ref{Jacobi}.
The coordinates $\Omega_\alpha=\{\phi_\alpha,\vartheta_\alpha,\varphi_\alpha\}$ determine the orientation of the plain containing the particles.
They are defined as follows: let $xyz$ be some laboratory system of coordinates and $x'''y'''z'''$ the body-fixed system of coordinates in which $\bm{y}_\alpha$ is along the $z'''$-axis and $\bm{x}_\alpha$ lies in the $x'''z'''$-plane. They are depicted in Fig.~\ref{Angles}~(d).
Then $\phi_\alpha,\vartheta_\alpha,\varphi_\alpha$ are the Euler angles of rotation of the laboratory to the body-fixed system of coordinates.
Three rotations in counter-clockwise direction are being done: the rotation around the $z$-axis by an angle $\phi_\alpha\in[0,2\pi)$ is followed by the rotation around the new $y'$-axis by an angle $\vartheta_\alpha\in[0,\pi)$, and the last rotation is around the new $z''$-axis by an angle $\varphi_\alpha\in[0,2\pi)$ (Fig.~\ref{Angles}).
These rotations are described by the standard rotation matrix~\cite{Mess61}
\begin{equation}
\scriptstyle
\mathfrak{R}(\phi_\alpha,\vartheta_\alpha,\varphi_\alpha)=
\left(
\begin{array}{ccc}
\cos\phi_\alpha\cos\vartheta_\alpha\cos\varphi_\alpha-\sin\phi_\alpha\sin\varphi_\alpha & -\cos\phi_\alpha\cos\vartheta_\alpha\sin\varphi_\alpha-\sin\phi_\alpha\cos\varphi_\alpha & \cos\phi_\alpha\sin\vartheta_\alpha \\
\sin\phi_\alpha\cos\vartheta_\alpha\cos\varphi_\alpha+\cos\phi_\alpha\sin\varphi_\alpha & -\sin\phi_\alpha\cos\vartheta_\alpha\sin\varphi_\alpha+\cos\phi_\alpha\cos\varphi_\alpha & \sin\phi_\alpha\sin\vartheta_\alpha \\
 -\sin\vartheta_\alpha\cos\varphi_\alpha & \sin\vartheta_\alpha\sin\varphi_\alpha & \cos\vartheta_\alpha
\end{array}
\right).
\end{equation}
Now the connection between sets of coordinates ($\bm{x}_\alpha$,$\bm{y}_\alpha$) and ($X_\alpha$,$\Omega_\alpha$) is easily expressed by
\begin{eqnarray}
\bm{x}_\alpha & = & x_\alpha\mathfrak{R}(\phi_\alpha,\vartheta_\alpha,\varphi_\alpha)
\left(
\begin{array}{c}
\sin\theta_\alpha\\
0\\
\cos\theta_\alpha
\end{array}
\right), \quad
\bm{y}_\alpha = y_\alpha\mathfrak{R}(\phi_\alpha,\vartheta_\alpha,\varphi_\alpha)
\left(
\begin{array}{c}
0\\
0\\
1
\end{array}
\right), \\
\label{jac-euler}
\tan\phi_\alpha & = & \frac{(\bm{y}_\alpha)_2}{(\bm{y}_\alpha)_1},\quad
\cos\vartheta_\alpha = \frac{(\bm{y}_\alpha)_3}{y_\alpha},\quad
\cos\varphi_\alpha = \frac{- (\bm{x}_\alpha)_3/x_\alpha+\cos\vartheta_\alpha\cos\theta_\alpha}{\sin\vartheta_\alpha\sin\theta_\alpha}.
\end{eqnarray}
In the case of formulae~(\ref{jac-euler}) the ambiguity of angle values is resolved by the fact that all the rotations described by $\mathfrak{R}(\phi_\alpha,\vartheta_\alpha,\varphi_\alpha)$ are done in counter-clockwise direction.
The connection between different sets of coordinates $X_\alpha$ and $X_\beta$ easily follows from~(\ref{JacobiTrans}):
\begin{eqnarray}
x_\beta & = & \sqrt{c_{\beta\alpha}^2 x_\alpha^2+2s_{\beta\alpha}c_{\beta\alpha}x_\alpha y_\alpha z_\alpha + s_{\beta\alpha}^2 y_\alpha^2}, \nonumber \\
y_\beta & = & \sqrt{s_{\beta\alpha}^2 x_\alpha^2-2s_{\beta\alpha}c_{\beta\alpha}x_\alpha y_\alpha z_\alpha + c_{\beta\alpha}^2 y_\alpha^2}, \nonumber \\
z_\beta & = & \frac{s_{\beta\alpha}c_{\beta\alpha}(y_\alpha^2-x_\alpha^2)+(c_{\beta\alpha}^2-s_{\beta\alpha}^2)x_\alpha y_\alpha z_\alpha}{x_\beta y_\beta}.
\end{eqnarray}
The connection between sets of Euler angles can be written in the form
\begin{equation}
\label{euler-ba}
\mathfrak{R}(\phi_\beta,\vartheta_\beta,\varphi_\beta) = \mathfrak{R}(\phi_\alpha,\vartheta_\alpha,\varphi_\alpha)\mathfrak{R}(0,w_{\beta\alpha},0),
\end{equation}
where $w_{\beta\alpha}$ is the angle such that the rotation by this angle in the plain containing particles places vector $y_\beta$ in the position of vector $y_\alpha$.
The equation~(\ref{euler-ba}) follows immediately from the identity $\mathfrak{R}^{-1}(0,w_{\beta\alpha},0)\mathfrak{R}^{-1}(\phi_\alpha,\vartheta_\alpha,\varphi_\alpha) = \mathfrak{R}^{-1}(\phi_\beta,\vartheta_\beta,\varphi_\beta)$ which simply expresses the fact that coordinates of a vector in the body-fixed frames defined with respect to different Jacobi coordinate sets $\alpha$ and $\beta$ are connected by the rotation matrix $\mathfrak{R}(0,w_{\beta\alpha},0)$.
With the orientation of Jacobi vectors chosen in this article (see Fig.~\ref{Jacobi}) the angle $w_{\beta\alpha}$ is given by
\begin{equation}
w_{\beta\alpha}=
\left\{
\begin{array}{l}
\arccos\frac{-s_{\beta\alpha}x_\alpha z_\alpha+c_{\beta\alpha}y_\alpha}{y_\beta},\quad \mathrm{if}\ (\beta,\alpha)=(1,2),(2,3),(3,1),\\
2\pi-\arccos\frac{-s_{\beta\alpha}x_\alpha z_\alpha+c_{\beta\alpha}y_\alpha}{y_\beta},\quad \mathrm{otherwise},
\end{array}
\right.
\end{equation}
where the range of $\arccos$ is $[0,\pi]$.

The change of variables in equations~(\ref{FM-6d}) can be derived quite easily with any symbolic computing system like Mathematica~\cite{Mathematica9}.
The equations now read
\begin{eqnarray}
\label{FM-6d-new}
\{ T_\alpha  +  V_\alpha(x_\alpha) +
\sum_{\beta\ne\alpha}V_\beta^{(\mathrm{l})}(x_\beta, y_\beta) &-& E \} \psi_\alpha(X_\alpha, \Omega_\alpha) = \nonumber \\
  &-&  V_\alpha^{(\mathrm{s})}(x_\alpha,y_\alpha) \sum_{\beta\ne\alpha}\psi_\beta(X_\beta, \Omega_\beta),\quad\alpha=1,2,3,
\end{eqnarray}
where $\psi_\alpha(X_\alpha, \Omega_\alpha)$ denote the components expressed in new variables.
The kinetic energy operator in new variables takes the form  
\begin{eqnarray}
T_\alpha & = & -\frac{1}{y_\alpha^2}\frac{\partial}{\partial y_\alpha}y_\alpha^2\frac{\partial}{\partial y_\alpha}
-\frac{1}{x_\alpha^2}\frac{\partial}{\partial x_\alpha}x_\alpha^2\frac{\partial}{\partial x_\alpha} \nonumber \\
 & - & \left(\frac{1}{y_\alpha^2}+\frac{1}{x_\alpha^2}\right)
\left(
\frac{1}{\sin\theta_\alpha}\frac{\partial}{\partial\theta_\alpha}\sin\theta_\alpha\frac{\partial}{\partial\theta_\alpha}+
\frac{1}{\sin^2\theta_\alpha}\frac{\partial^2}{\partial\varphi_\alpha^2}
\right)+
\frac{\bm{J}^2-\bm{K}_\alpha}{y_\alpha^2},
\end{eqnarray}
where  $\bm{J}^2$ is the total 
orbital  momentum squared operator
\begin{equation}
\bm{J}^2 = -\left[
\frac{1}{\sin\vartheta_\alpha}\frac{\partial}{\partial\vartheta_\alpha}\sin\vartheta_\alpha\frac{\partial}{\partial\vartheta_\alpha}+
\frac{1}{\sin^2\vartheta_\alpha}\left(
\frac{\partial^2}{\partial\phi_\alpha^2}-2\cos\vartheta_\alpha\frac{\partial^2}{\partial\phi_\alpha\partial\varphi_\alpha}+\frac{\partial^2}{\partial\varphi_\alpha^2}
\right)
\right]
\end{equation}
and $ \bm{K}_\alpha$ is given by 
\begin{equation}
\bm{K}_\alpha = \frac{\partial}{\partial\theta_\alpha}
\left(
\bm{J}_\alpha^{(+)}+\bm{J}_\alpha^{(-)}
\right)
+\cot\theta_\alpha
\left(
\bm{J}_\alpha^{(+)}-\bm{J}_\alpha^{(-)}
\right)
+2\bm{J}_{z'}^2
\end{equation}
with
\begin{equation}
\bm{J}_\alpha^{(\pm)} = \mp\mathrm{e}^{\mp\mathrm{i}\varphi_\alpha}\left[\pm\frac{\partial}{\partial\vartheta_\alpha}+
\frac{\mathrm{i}}{\sin\vartheta_\alpha}\frac{\partial}{\partial\phi_\alpha}-
\mathrm{i}\cot\vartheta_\alpha\frac{\partial}{\partial\varphi_\alpha}\right],\quad
\bm{J}_{z'} = -\mathrm{i}\partial/\partial\varphi_\alpha.
\end{equation}

Now we introduce Wigner $D$-functions $D_{M,M'}^{J}$
\begin{equation}
\label{D-def}
D_{MM'}^J(\phi_\alpha,\vartheta_\alpha,\varphi_\alpha) =
\mathrm{e}^{-\mathrm{i}M\phi_\alpha}d_{MM'}^J(\vartheta_\alpha)\mathrm{e}^{-\mathrm{i}M'\varphi_\alpha},
\end{equation}
where
\begin{equation}
d_{MM'}^J(\vartheta_\alpha) = 
\sqrt{\frac{(J+M')!(J-M)!}{(J+M)!(J-M)!}}
\left(\sin\frac{\vartheta_\alpha}{2}\right)^{M'-M}\left(\cos\frac{\vartheta_\alpha}{2}\right)^{M'+M}
P_{J-M'}^{(M'-M,\,M'+M)}(\cos\vartheta_\alpha).
\end{equation}
Here $P_n^{(\alpha,\beta)}$ are the Jacobi 
polynomials~\cite{NISTDLMF}.
The definition~(\ref{D-def}) coincides with those of~\cite{Bieden81,Varsh89}, but differs from that of~\cite{Kostr89}.
Wigner $D$-functions are the common eigenfunctions of total 
orbital momentum squared $\bm{J}^2$ and its projection $\bm{J}_z = -\mathrm{i}\partial/\partial\phi_\alpha$ operators
\begin{equation}
\label{D-eigen}
\bm{J}^2 D_{MM'}^J = J(J+1)D_{MM'}^J,\quad
\bm{J}_zD_{MM'}^J = -M D_{MM'}^J.
\end{equation}
They also obey
\begin{equation}
\label{D-pm}
\bm{J}_\alpha^{(\pm)} D_{MM'}^J = 
\pm\lambda^{J,\pm M'} D_{MM'\pm1}^J,\quad
\bm{J}_{z'} D_{MM'}^J = -M'D_{MM'}^J,
\end{equation}
where
\begin{equation}
\lambda^{JM'}=\sqrt{J(J+1)-M'(M'+1)}.
\end{equation}
The orthogonality conditions are
\begin{equation}
\label{D-ortho}
\int_0^{2\pi}\mathrm{d}\phi_\alpha\,
\int_0^{2\pi}\mathrm{d}\varphi_\alpha\,
\int_0^{\pi}\mathrm{d}\vartheta_\alpha\,
\sin\vartheta_\alpha
\left(D_{M_1M'_1}^{J_1}(\phi_\alpha,\vartheta_\alpha,\varphi_\alpha)\right)^*
D_{M_2M'_2}^{J_2}(\phi_\alpha,\vartheta_\alpha,\varphi_\alpha) =
\frac{8\pi^2}{2J_1+1}\delta_{J_1J_2}\delta_{M_1M_2}\delta_{M'_1M'_2}.
\end{equation}
Due to the rotation transformation~(\ref{euler-ba}) the Wigner $D$-functions of arguments with different index $\alpha$ are bound by the following relationship
\begin{equation}
\label{D-ba}
D_{MM'}^J(\phi_\beta,\vartheta_\beta,\varphi_\beta)=
\sum_{M''=-J}^{J}D_{MM''}^J(\phi_\alpha,\vartheta_\alpha,\varphi_\alpha)D_{M''M'}^J(0,w_{\beta\alpha},0).
\end{equation}
Wigner $D$-functions with integer $J\ge0$ and $|M|,|M'|\le J$ form a basis in the space of square integrable functions on the domain~$[0,2\pi]\times[0,\pi]\times[0,2\pi]$.
However, as a basis for the expansion of FM components we use linear combinations of Wigner $D$-functions defined as
\begin{equation}
\label{F-def}
F_{MM'}^{J\tau}(\Omega_\alpha) =
\frac{1}{\sqrt{2+2\delta_{M'0}}}\left( D_{MM'}^J(\Omega_\alpha)+\tau(-1)^{M'}D_{M,-M'}^J(\Omega_\alpha) \right),
\end{equation}
where $\tau=\pm1$ is the parity.
They are constructed to be the eigenfunctions not only 
of $\bm{J}^2$ and $\bm{J}_z$ but also of the inversion operator
\begin{equation}
\bm{P}F_{MM'}^{J\tau}(\phi_\alpha,\vartheta_\alpha,\varphi_\alpha)= F_{MM'}^{J\tau}(\phi_\alpha+\pi,\pi-\vartheta_\alpha,\pi-\varphi_\alpha) = \tau(-1)^J F_{MM'}^{J\tau}(\phi_\alpha,\vartheta_\alpha,\varphi_\alpha).
\end{equation}
This equation follows easily from the identity $D_{MM'}^J(\phi_\alpha+\pi,\pi-\vartheta_\alpha,\pi-\varphi_\alpha)=(-1)^{J+M'}D_{M,-M'}^J(\phi_\alpha,\vartheta_\alpha,\varphi_\alpha)$.
Equations~(\ref{D-eigen}),~(\ref{D-pm}) for $F_{MM'}^{J\tau}$ become
\begin{eqnarray}
\label{F-eigen}
\bm{J}^2 F_{MM'}^{J\tau} & = & J(J+1)F_{MM'}^{J\tau},\quad
\bm{J}_zF_{MM'}^{J\tau} = -M F_{MM'}^{J\tau},\quad
\bm{J}_{z'} F_{MM'}^{J\tau} = -M' F_{MM'}^{J,-\tau}, \nonumber \\
\left(\bm{J}_\alpha^{(+)}+\bm{J}_\alpha^{(-)}\right) F_{MM'}^{J\tau} & = &
\lambda^{J,M'} F_{M,M'+1}^{J\tau}\sqrt{1+\delta_{M'0}}(1-\delta_{M'0}\delta_{\tau,-1}) \nonumber \\
 & - & \lambda^{J,-M'}F_{M,M'-1}^{J\tau}\sqrt{1+\delta_{M'1}}(1-\delta_{M'0})(1-\delta_{M'1}\delta_{\tau,-1}), \nonumber \\
\left(\bm{J}_\alpha^{(+)}-\bm{J}_\alpha^{(-)}\right)\bm{J}_{z'} F_{MM'}^{J\tau} & = &
-M'\lambda^{J,M'} F_{M,M'+1}^{J\tau}\sqrt{1+\delta_{M'0}}(1-\delta_{M'0}\delta_{\tau,-1}) \nonumber \\
 & - & M'\lambda^{J,-M'}F_{M,M'-1}^{J\tau}\sqrt{1+\delta_{M'1}}(1-\delta_{M'0})(1-\delta_{M'1}\delta_{\tau,-1}).
\end{eqnarray}
The orthogonality conditions are easily deduced from~(\ref{D-ortho})
\begin{equation}
\label{F-ortho}
\int_0^{2\pi}\mathrm{d}\phi_\alpha\,
\int_0^{2\pi}\mathrm{d}\varphi_\alpha\,
\int_0^{\pi}\mathrm{d}\vartheta_\alpha\,
\sin\vartheta_\alpha
\left(F_{M_1M'_1}^{J_1\tau_1}(\phi_\alpha,\vartheta_\alpha,\varphi_\alpha)\right)^*
F_{M_2M'_2}^{J_2\tau_2}(\phi_\alpha,\vartheta_\alpha,\varphi_\alpha) =
\frac{8\pi^2}{2J_1+1}\delta_{J_1J_2}\delta_{\tau_1\tau_2}\delta_{M_1M_2}\delta_{M'_1M'_2}
\end{equation}
and the
relation between the functions of different arguments follows from~(\ref{D-ba})
\begin{equation}
\label{F-ba}
F_{MM'}^{J\tau}(\phi_\beta,\vartheta_\beta,\varphi_\beta)=
\sum_{M''=M_0}^{J}(-1)^{M'-M''}\frac{2}{\sqrt{2+2\delta_{M'0}}}F_{M'M''}^{J\tau}(0,w_{\beta\alpha},0)F_{MM''}^{J\tau}(\phi_\alpha,\vartheta_\alpha,\varphi_\alpha),
\end{equation}
where we have defined
\begin{equation}
M_0 = (1-\tau)/2.
\end{equation}
Finally, from the definition~(\ref{F-def}) and the identity $F_{M,-M'}^{J\tau}=\tau(-1)^{M'}F_{M,M'}^{J\tau}$ it follows that the functions $F_{MM'}^{J\tau}$ with integer $J\ge0$, $\tau=\pm1$, $|M|\le J$ and $(1-\tau)/2\le M'\le J$ also form a basis in the space of square integrable functions on the domain~$[0,2\pi]\times[0,\pi]\times[0,2\pi]$.

We are now ready to expand the FM components in terms of the functions $F_{MM'}^{J\tau}$
\begin{equation}
\label{exp-full}
\psi_\alpha(X_\alpha,\Omega_\alpha) = 
\sum_{J=0}^{+\infty}\sum_{\tau=\pm1}\sum_{M=-J}^{J}\sum_{M'=M_0}^{J}
\frac{\psi_{\alpha MM'}^{J\tau}(X_\alpha)}{x_\alpha y_\alpha}
F_{MM'}^{J\tau}(\phi_\alpha,\vartheta_\alpha,\varphi_\alpha).
\end{equation}
The factor $1/(x_\alpha y_\alpha)$ is introduced here to get rid of the first derivatives in $x_\alpha$ and $y_\alpha$ in the resulting equations.
Substituting~(\ref{exp-full}) into the FM equations~(\ref{FM-6d}), projecting them onto the basis functions $F_{MM'}^{J\tau}$ and using equations~(\ref{F-eigen})--(\ref{F-ba}) we obtain the finite set of 3D equations for partial components~$\psi_{\alpha MM'}^{J\tau}(X_\alpha)$  
\begin{eqnarray}
\label{FM}
 & & \left[
T_{\alpha MM'}^{J\tau}+V_{\alpha}(x_\alpha)+
\sum_{\beta\ne\alpha}V_{\beta}^{(\mathrm{l})}(x_\beta,y_\beta)-E
\right]
\psi_{\alpha MM'}^{J\tau}(X_\alpha) \nonumber \\
 & + &
T_{\alpha M,M'-1}^{J\tau-} \psi_{\alpha M,M'-1}^{J\tau}(X_\alpha)+
T_{\alpha M,M'+1}^{J\tau+} \psi_{\alpha M,M'+1}^{J\tau}(X_\alpha) \nonumber \\
 & = & -V_{\alpha}^{(\mathrm{s})}(x_\alpha,y_\alpha)
\sum_{\beta\ne\alpha}\frac{x_\alpha y_\alpha}{x_\beta y_\beta}
\sum_{M''=M_0}^{J}(-1)^{M''-M'}
\frac{2}{\sqrt{2+2\delta_{M''0}}}F_{M''M'}^{J\tau}(0,w_{\beta\alpha},0)
\psi_{\beta MM''}^{J\tau}(X_\beta).
\end{eqnarray}
Here the kinetic part is of the form 
\begin{equation}
T_{\alpha MM'}^{J\tau} = -\frac{\partial^2}{\partial y_\alpha^2} +
\frac{1}{y_\alpha^2}\left( J(J+1) - 2M'^2 \right) - 
\frac{\partial^2}{\partial x_\alpha^2} -
\left( \frac{1}{y_\alpha^2}+\frac{1}{x_\alpha^2} \right)
\left(
\frac{\partial}{\partial z_\alpha}(1-z_\alpha^2)\frac{\partial}{\partial z_\alpha} -
\frac{M'^2}{1-z_\alpha^2}
\right),
\end{equation}
\begin{equation}
T_{\alpha M,M'\pm1}^{J\tau\pm} =
\pm\frac{1}{y_\alpha^2}\lambda^{J,\pm M'}
\sqrt{1+\delta_{M'0(1)}}
\left[
-\sqrt{1-z_\alpha^2}\frac{\partial}{\partial z_\alpha}\pm
(M'\pm 1)\frac{z_\alpha}{\sqrt{1-z_\alpha^2}}
\right].
\end{equation}
The key property of the obtained equations is that equations with different $J$, $\tau$ and $M$ 
form independent sets of equations.
This is the direct consequence of the fact that the total
orbital momentum, its projection and the parity are the integrals of motion for the three-particle system.
For given $J$, $\tau$ and $M$ the equations~(\ref{FM}) are enumerated by indices $M'=M_0,\ldots,J$ and $\alpha = 1,2,3$ thus forming a finite set of $3n_M$ three dimensional PDEs.
Here $n_M = J-M_0+1$ is the number of possible values of $M'$ for given $J$ and $\tau$.
In addition to zero Dirichlet asymptotic boundary conditions
the partial components $\psi_{\alpha MM'}^{J\tau}$ must also satisfy zero boundary conditions on the lines $x_\alpha=0$, $y_\alpha = 0$ to be continuous.

Equations~(\ref{FM}) are called the (3D) FM equations in total 
orbital momentum representation. For the first time they were obtained in~\cite{Kostr89} in a slightly different form.
They are the main result
of this section.

\subsection{Permutational symmetry}

As is well known from the basics of quantum mechanics, presence of identical particles in the system leads to a symmetry (antisymmetry) of the wave function.
Here we show how it can be used to reduce the number of 3D FM equations.
In the three-particle system two cases of two or three identical particles are possible.
Let the particles $\alpha$ and $\beta$ be identical.
Then the operator $P_{\alpha\beta}$ which permutes the coordinates of particles $\alpha$ and $\beta$ in a wave function is an additional integral of motion for a three-particle system and commutes with the full Hamiltonian of a system.
As a result, equations for symmetric and antisymmetric parts of the wave function can be solved independently.

Let $\psi_{\alpha}$ be the FM components of a symmetric or antisymmetric part of the wave function $\Psi=\psi_1+\psi_2+\psi_3$ satisfying $P_{\alpha\beta}\Psi = p\Psi$ with $p=1$ or $p = -1$ respectively.
We now find the constraints on the components in these cases.
Consider for definiteness the symmetry (antisymmetry) with respect to operator $P_{12}$.
Note that permutation of particles 1 and 2 leads to the following change in pairs of Jacobi vectors:
\begin{equation}
(\bm{x}_1, \bm{y}_1) \to (-\bm{x}_2, \bm{y}_2),\quad
(\bm{x}_2, \bm{y}_2) \to (-\bm{x}_1, \bm{y}_1),\quad
(\bm{x}_3, \bm{y}_3) \to (-\bm{x}_3, \bm{y}_3).
\end{equation}
Then from the relation
\begin{eqnarray}
P_{12}\big( \psi_1(\bm{x}_1, \bm{y}_1)+\psi_2(\bm{x}_2, \bm{y}_2)+\psi_3(\bm{x}_3, \bm{y}_3) \big) & = & \psi_1(-\bm{x}_2, \bm{y}_2)+\psi_2(-\bm{x}_1, \bm{y}_1)+\psi_3(-\bm{x}_3, \bm{y}_3) \nonumber \\
& = & p\psi_1(\bm{x}_1, \bm{y}_1)+p\psi_2(\bm{x}_2, \bm{y}_2)+p\psi_3(\bm{x}_3, \bm{y}_3)
\end{eqnarray}
we obtain the constraints
\begin{equation}
\label{2-sym}
\psi_2(\bm{x}_2, \bm{y}_2) = p\psi_1(-\bm{x}_2, \bm{y}_2),\quad \psi_3(\bm{x}_3, \bm{y}_3) = p\psi_3(-\bm{x}_3, \bm{y}_3).
\end{equation}
Thus in the case of two identical particles 1 and 2 the component $\psi_3$ is symmetric (antisymmetric) with respect to its argument $\bm{x}_3$ and components $\psi_1$ and $\psi_2$ are simply related by the first identity of~(\ref{2-sym}).
In the case of three identical particles and fully (anti)symmetric wave function it obeys $P_{12}\Psi=P_{23}\Psi=P_{13}\Psi=p\Psi$, then additionally we have
\begin{equation}
\label{3-sym}
\psi_1(\bm{x}_1, \bm{y}_1)=\psi_2(\bm{x}_1, \bm{y}_1)=\psi_3(\bm{x}_1, \bm{y}_1)=p\psi_1(-\bm{x}_1, \bm{y}_1).
\end{equation}
In coordinates $(X_\alpha,\Omega_\alpha)$ the substitution $(\bm{x}_\alpha, \bm{y}_\alpha)\to (-\bm{x}_\alpha, \bm{y}_\alpha)$ reads
\begin{equation}
(x_\alpha, y_\alpha, z_\alpha, \phi_\alpha, \vartheta_\alpha, \varphi_\alpha) \to
\left(x_\alpha, y_\alpha, -z_\alpha, \phi_\alpha, \vartheta_\alpha,
\left\{
\begin{array}{l}
\varphi_\alpha + \pi, \varphi_\alpha \in [0, \pi), \\
\varphi_\alpha - \pi, \varphi_\alpha \in [\pi, 2\pi)
\end{array}
\right.
\right).
\end{equation}
Then from the identity $F_{MM'}^{J\tau}(\phi_\alpha,\vartheta_\alpha,\varphi_\alpha\pm\pi) = (-1)^{M'}F_{MM'}^{J\tau}(\phi_\alpha,\vartheta_\alpha,\varphi_\alpha)$ and the definition of partial components~(\ref{exp-full}) we get
\begin{equation}
\psi_\alpha(-\bm{x}_\alpha, \bm{y}_\alpha) = \sum_{J\tau MM'}(-1)^{M'}\frac{\psi_{\alpha MM'}^{J\tau}(x_\alpha, y_\alpha, -z_\alpha)}{x_\alpha y_\alpha}F_{MM'}^{J\tau}(\phi_\alpha,\vartheta_\alpha,\varphi_\alpha).
\end{equation}
Thus, finally, the constraints~(\ref{2-sym}) and~(\ref{3-sym}) for partial components take the form
\begin{equation}
\label{2-sym-part}
\psi_{2MM'}^{J\tau}(x_2, y_2, z_2) = p(-1)^{M'}\psi_{1MM'}^{J\tau}(x_2, y_2, -z_2),\quad
\psi_{3MM'}^{J\tau}(x_3, y_3, z_3) = p(-1)^{M'}\psi_{3MM'}^{J\tau}(x_3, y_3, -z_3)
\end{equation}
and
\begin{equation}
\label{3-sym-part}
\psi_{1MM'}^{J\tau}(x_1,y_1,z_1)=\psi_{2MM'}^{J\tau}(x_1,y_1,z_1)=\psi_{3MM'}^{J\tau}(x_1,y_1,z_1)=p(-1)^{M'}\psi_{1MM'}^{J\tau}(x_1,y_1,-z_1).
\end{equation}
The relations~(\ref{2-sym-part}) and~(\ref{3-sym-part}) allow one to solve equations~(\ref{FM}) only for components $\psi_{\alpha MM'}^{J\tau}$ with $\alpha=1,3$ in the case of two identical particles and with $\alpha=1$ in the case of three identical particles.
The number of coupled three-dimensional equations is then reduced from $3n_M$ to $2n_M$ and $n_M$, respectively.
Moreover, the (anti)symmetry of functions $\psi_{3MM'}^{J\tau}$ and $\psi_{1MM'}^{J\tau}$ in~(\ref{2-sym-part}) and~(\ref{3-sym-part}) reduces the $z_\alpha$-domain in corresponding equations to $z_\alpha\in[0,1]$.

\section{The computational scheme}

\subsection{Basic scheme}

We start this section by transforming the equations~(\ref{FM}) by making a substitution
\begin{equation}
\label{z-subst}
\psi_{\alpha MM'}^{J\tau}(x_\alpha,y_\alpha,z_\alpha) = 
( 1-z_\alpha^2 )^{\frac{M'}{2}}
\widehat{\psi}_{\alpha MM'}^{J\tau}(x_\alpha,y_\alpha,z_\alpha).
\end{equation}
The reason for that is as follows.
It can be shown that the singular term $M'^2/(1-z_\alpha^2)$ in operator $T_{\alpha MM'}^{J\tau}$
enforces the partial components to have the following behaviour at $z_\alpha=\pm1$
\begin{equation}
\left.
\psi_{\alpha MM'}^{J\tau}(x_\alpha,y_\alpha,z_\alpha)
\right|_{z_\alpha\to\pm1}\sim
( 1-z_\alpha^2 )^{\frac{M'}{2}}.
\end{equation}
It makes problematic the use of smooth bases for the expansion of partial components with odd $M'$.
Substitution~(\ref{z-subst}) corrects it.
The new partial components $\widehat{\psi}_{\alpha MM'}^{J\tau}$ obey the equations
\begin{eqnarray}
\label{FM-pre}
 & & \left[
\widehat{T}_{\alpha MM'}^{J\tau}+V_{\alpha}(x_\alpha)+
\sum_{\beta\ne\alpha}V_{\beta}^{(\mathrm{l})}(x_\beta,y_\beta)-E
\right]
\widehat{\psi}_{\alpha MM'}^{J\tau}(X_\alpha) \nonumber \\
 & + & \widehat{T}_{\alpha M,M'-1}^{J\tau-} \widehat{\psi}_{\alpha M,M'-1}^{J\tau}(X_\alpha)+
\widehat{T}_{\alpha M,M'+1}^{J\tau+} \widehat{\psi}_{\alpha M,M'+1}^{J\tau}(X_\alpha) \nonumber \\
 & = & -\frac{V_{\alpha}^{(\mathrm{s})}(x_\alpha,y_\alpha)}{(1-z_\alpha^2)^{\frac{M'}{2}}}
\sum_{\beta\ne\alpha}\frac{x_\alpha y_\alpha}{x_\beta y_\beta}
\sum_{M''=M_0}^{J}(-1)^{M''-M'}
\frac{2}{\sqrt{2+2\delta_{M''0}}}F_{M''M'}^{J\tau}(0,w_{\beta\alpha},0)
(1-z_\beta^2)^{\frac{M''}{2}}
\widehat{\psi}_{\beta MM''}^{J\tau}(X_\beta) \nonumber \\
 & & 
\end{eqnarray}
where the new kinetic part operators are
\begin{eqnarray}
\widehat{T}_{\alpha MM'}^{J\tau} & = & -\frac{\partial^2}{\partial y_\alpha^2} +
\frac{1}{y_\alpha^2}\left( J(J+1) - 2M'^2 \right) - 
\frac{\partial^2}{\partial x_\alpha^2} \nonumber \\
 & - & \left( \frac{1}{y_\alpha^2}+\frac{1}{x_\alpha^2} \right)
\left(
(1-z_\alpha^2)\frac{\partial^2}{\partial z_\alpha^2} -
2(M'+1)z_\alpha\frac{\partial}{\partial z_\alpha} -
M'(M'+1)
\right),
\end{eqnarray}
\begin{eqnarray}
\widehat{T}_{M,M'+1}^{J\tau+}  & = &
\frac{1}{y_\alpha^2}
\widehat{\lambda}^{JM'}
\left[
-(1-z_\alpha^2)\frac{\partial}{\partial z_\alpha}+
2(M'+1)z_\alpha
\right], \nonumber \\
\widehat{T}_{M,M'-1}^{J\tau-}  & = &
\frac{1}{y_\alpha^2}
\widehat{\lambda}^{J,-M'}
\frac{\partial}{\partial z_\alpha},
\end{eqnarray}
where we have introduced
\begin{equation}
\widehat{\lambda}^{J,\pm M'}=\lambda^{J,\pm M'}\sqrt{1+\delta_{M'0(1)}}.
\end{equation}
Now we make the last but not least step of preparing the 3D FM equations for discretization.
For that we represent the sum of the tail parts of potentials at the left hand side of equations in the form
\begin{equation}
\sum_{\beta\ne\alpha}V_{\beta}^{(\mathrm{l})}(x_\beta,y_\beta) =
\widehat{V}_\alpha(y_\alpha) +
\left(
\sum_{\beta\ne\alpha}V_{\beta}^{(\mathrm{l})}(x_\beta,y_\beta) - \widehat{V}_\alpha(y_\alpha)
\right)
\end{equation}
and rearrange terms in equations so that they become
\begin{eqnarray}
\label{FM-fin}
 & & \left[
\widehat{T}_{\alpha MM'}^{J\tau}+V_{\alpha}(x_\alpha)+
\widehat{V}_{\alpha}(y_\alpha)-E
\right]
\widehat{\psi}_{\alpha MM'}^{J\tau}(X_\alpha)+
\widehat{T}_{\alpha M,M'-1}^{J\tau-} \widehat{\psi}_{\alpha M,M'-1}^{J\tau}(X_\alpha)+
\widehat{T}_{\alpha M,M'+1}^{J\tau+} \widehat{\psi}_{\alpha M,M'+1}^{J\tau}(X_\alpha) \nonumber \\
 & = & -\frac{V_{\alpha}^{\mathrm{s}}(x_\alpha,y_\alpha)}{(1-z_\alpha^2)^{\frac{M'}{2}}}
\sum_{\beta\ne\alpha}\frac{x_\alpha y_\alpha}{x_\beta y_\beta}
\sum_{M''=M_0}^{J}(-1)^{M''-M'}
\frac{2}{\sqrt{2+2\delta_{M''0}}}F_{M''M'}^{J\tau}(0,w_{\beta\alpha},0)
(1-z_\beta^2)^{\frac{M''}{2}}
\widehat{\psi}_{\beta MM''}^{J\tau}(X_\beta) \nonumber \\
 & - &
\left(
\sum_{\beta\ne\alpha}V_{\beta}^{(\mathrm{l})}(x_\beta,y_\beta) - \widehat{V}_\alpha(y_\alpha)
\right)
\widehat{\psi}_{\alpha M,M'}^{J\tau}(X_\alpha).
\end{eqnarray}
The potential $\widehat{V}_\alpha$ is an approximation of $\sum_{\beta\ne\alpha}V_{\beta}^{(\mathrm{l})}$ in the sense that the new right hand side term $(\sum_{\beta\ne\alpha}V_{\beta}^{(\mathrm{l})}-\widehat{V}_\alpha)\widehat{\psi}_{\alpha M,M'}^{J\tau}$ is small.
For better theoretical and computational properties of the resulting equations the potential $\widehat{V}_\alpha$'s choice must secure the square integrability of this term.
The actual form depends on the problem that is solved.
For bound states and below the breakup threshold scattering calculations the possible choice is
$\widehat{V}_\alpha(y_\alpha)=Z_\alpha(Z_\beta+Z_\gamma)\sqrt{2\mu_{\alpha(\beta\gamma)}}/(y_\alpha+1)$
(this is the leading term of $\sum_{\beta\ne\alpha}V_{\beta}^{(\mathrm{l})}$ when $y_\alpha\to+\infty$ with $x_\alpha$ fixed).
Equations~(\ref{FM-fin}) are the final form of 3D FM equations in total
orbital momentum representation that are discretized.
The reason for the rearrangement done is that now variables ``almost separate'' in the operator at the left hand side and it can be used to construct an effective preconditioner in the computational scheme.

In a nutshell, our scheme is based on the collocation method with local basis.
It results in the generalized eigenvalue problem of the form
\begin{equation}
H\bm{c} = ES\bm{c}.
\end{equation}
To find eigenvalues in the vicinity of a given $E=E^*$ it is rewritten in the form
\begin{equation}
\label{FM-lin}
S(H-E^*S)^{-1}\widetilde{\bm{c}}=\frac{1}{E-E^*}\widetilde{\bm{c}}
\end{equation}
with $\widetilde{\bm{c}} = (H-E^*S)\bm{c}$.
Then an iterative method is applied to find the eigenvalues of the matrix $S(H-E^*S)^{-1}$.
For solving systems of linear equations with the matrix $(H-E^*S)$ we use a preconditioning matrix which is an approximation of the inverse matrix of the operator at the left-hand side of equations~(\ref{FM-fin}) with $E=E^*$.
This choice of a preconditioner is exploited in~\cite{Schelling89, Lazau03, Roudn00}.

Let the functions $S_\alpha^i(x)$, $S_\alpha^j(y)$ and $S_\alpha^k(z)$ form bases of local functions defined on given intervals $[0,R_{x_\alpha}]$, $[0,R_{y_\alpha}]$ and $[-1,1]$ and $S_\alpha^i(x)$, $S_\alpha^j(y)$ satisfy zero Dirichlet-type boundary conditions at the endpoints.
Denote by $r$ the highest possible number of basis functions that are nonzero at any point of interval of definition for all sets $S_\alpha^i$, $S_\alpha^j$ and $S_\alpha^k$.
We seek the solutions $\widehat{\psi}_{\alpha MM'}^{J\tau}$ of~(\ref{FM-fin}) in the cubes $[0,R_{x_\alpha}]\times[0,R_{y_\alpha}]\times[-1,1]$ expanding them in terms of products of those basis functions
\begin{equation}
\widehat{\psi}_{\alpha MM'}^{J\tau}(x_\alpha,y_\alpha,z_\alpha) = 
\sum_{i,j,k=1}^{n_{x_\alpha},n_{y_\alpha},n_{z_\alpha}}
c_{ijk}^{M'}S^i_\alpha(x_\alpha)S^j_\alpha(y_\alpha)S^k_\alpha(z_\alpha).
\end{equation}
We use the same basis set for partial components $\widehat{\psi}_{\alpha MM'}^{J\tau}$ with different $M'$.
This fact is not used explicitly in constructing the scheme which can be easily generalized on the case of $M'$ dependent basis sets.

Let us introduce regular grids of collocation points $(x^\xi_\alpha,y^\eta_\alpha,z^\zeta_\alpha)$ with $\xi = 1,\ldots,n_{x_\alpha}$, $\eta = 1,\ldots,n_{y_\alpha}$, $\zeta = 1,\ldots,n_{z_\alpha}$.
Again for notation simplicity we use identical sets of collocation points for equations~(\ref{FM-fin}) with different $M'$.
Requiring the equations~(\ref{FM-fin}) to be satisfied at collocation points,
we obtain the generalized eigenvalue problem~(\ref{FM-lin}).
The collocation matrix $H$ is the discretized version of 3D FM equations~(\ref{FM-fin}) operator with $E=0$ and $S$ is the matrix with elements $S^i_\alpha(x^\xi_\alpha)S^j_\alpha(y^\eta_\alpha)S^k_\alpha(z^\zeta_\alpha)$.
The obtained matrix $H$ has linear dimension $n_M\sum_{\alpha}n_{x_\alpha}n_{y_\alpha}n_{z_\alpha}$.
However, due to the use of local bases the number of nonzero elements of $H$ is $O(n_M^2\sum_{\alpha}n_{x_\alpha}n_{y_\alpha}n_{z_\alpha})$ only.

\subsection{Preconditioner}

We now turn to construction of the preconditioner that allows us to efficiently invert the operator at the left-hand side of equations~(\ref{FM-fin}).
For that we use tensor product methods based on the technique that is known as Matrix Decomposition Algorithm (MDA)~\cite{Bial11} or Tensor Trick (TT)~\cite{Schelling89}.
Let us first outline some preliminary facts.
The Kronecker product of $n\times n$ matrix $A$ and $m\times m$ matrix $B$ is the $(nm)\times (nm)$ matrix $A\otimes B$ with elements $(A\otimes B)_{(i-1)m+k,(j-1)m+l} = a_{ij}b_{kl}$. The following properties are satisfied:
\begin{enumerate}
\item
\begin{equation}
(A_1\otimes B_1)(A_2\otimes B_2) = (A_1A_2)\otimes(B_1B_2).
\end{equation}
\item
\begin{equation}
(A\otimes B)^{-1} = A^{-1}\otimes B^{-1}.
\end{equation}
\end{enumerate}
MDA (TT) is based on the folowing trick.
Suppose we need to invert the $(nm)\times(nm)$ matrix of the form
\begin{equation}
A_1\otimes B_1 + A_2\otimes B_2
\end{equation}
with $n\times n$ matrices $A_{1,2}$ and $m\times m$ matrices $B_{1,2}$.
Let now $\overline{W}_A$, $W_A$ and $\Lambda_A$ be (if exist) nonsingular complex matrices which are the solutions of left and right generalized eigenvalue problems
\begin{equation}
\overline{W}_A A_1 = \Lambda_A\overline{W}_A A_2,\quad A_1 W_A = A_2 W_A \Lambda_A.
\end{equation}
Then the matrices $\overline{W}_A$ and $W_A$ simultaneously diagonalize matrices $A_1$ and $A_2$.
They can be scaled to satisfy
\begin{equation}
\label{WW-scale}
\overline{W}_A A_1 W_A = \Lambda_A,\quad \overline{W}_A A_2 W_A = I,
\end{equation}
where $I$ is the identity matrix of the proper size.
In the following we denote the operation of finding the solutions of generalized eigenvalue problems that satisfy~(\ref{WW-scale}) by $\mathfrak{D}$
\begin{equation}
(\overline{W}_A,W_A,\Lambda_A) = \mathfrak{D}(A_1,A_2).
\end{equation}
We note that the computational cost of $\mathfrak{D}$ is $O(n^3)$.
Now let also
\begin{equation}
(\overline{W}_B,W_B,\Lambda_B) = \mathfrak{D}(B_1,B_2).
\end{equation}
Then one obtains
\begin{equation}
A_1\otimes B_1 + A_2\otimes B_2 =
(\overline{W}_A^{\,-1}\otimes \overline{W}_B^{\,-1})(\Lambda_A\otimes \Lambda_B+I\otimes I)(W_A^{-1}\otimes W_B^{-1})
\end{equation}
and consequently
\begin{equation}
\label{MDA}
(A_1\otimes B_1 + A_2\otimes B_2)^{-1} =
(W_A\otimes W_B)(\Lambda_A\otimes \Lambda_B+I\otimes I)^{-1}(\overline{W}_A\otimes \overline{W}_B).
\end{equation}
The matrix $\Lambda_A\otimes \Lambda_B+I\otimes I$ is diagonal and is inverted fast.
With the use of~(\ref{MDA}) the inversion of the matrix requires $O(n^3+m^3)$ instead of $O(n^3m^3)$ operations if treating it as a matrix of general form.
The matrix-vector product with the matrix of the form~(\ref{MDA}) requires $O(nm(n+m))$ multiplications instead of the ordinary $O(n^2m^2)$.
The review of theory and applications of MDAs can be found in~\cite{Bial11}.

We denote the discretized version of the operator at the left-hand side of equations~(\ref{FM-fin}) with $E=E^*$ by $L$.
This matrix is block diagonal with blocks $L_\alpha$ that can be inverted independently.
Consider the $L_\alpha$ block.
It is block tridiagonal with blocks $L_{\alpha M',M'-1}$, $L_{\alpha M'M'}$ and $L_{\alpha M',M'+1}$ in row $M' = M_0,\ldots,J$.
Let the collocation points $(x^\xi_\alpha,y^\eta_\alpha,z^\zeta_\alpha)$ and three-dimensional basis functions $S^i_\alpha S^j_\alpha S^k_\alpha$ are enumerated so that the most frequently changing index is $\eta$, then $\xi$ and then $\zeta$ (and so as $j,i,k$).
Then the blocks are of the form
\begin{eqnarray}
L_{\alpha M'M'} & = & -S_{z_\alpha}\otimes S_{x_\alpha}\otimes D_{y_\alpha}
- S_{z_\alpha}\otimes D_{x_\alpha}\otimes S_{y_\alpha}
+ \left( J(J+1)-2M'^2 \right) S_{z_\alpha}\otimes S_{x_\alpha}\otimes (Y_{\alpha r2}S_{y_\alpha}) \nonumber \\
 & - & D_{z_\alpha}^{M'}\otimes S_{x_\alpha}\otimes (Y_{\alpha r2}S_{y_\alpha})
-D_{z_\alpha}^{M'}\otimes (X_{\alpha r2}S_{x_\alpha})\otimes S_{y_\alpha} \nonumber \\
 & + & S_{z_\alpha}\otimes (V_\alpha S_{x_\alpha})\otimes S_{y_\alpha}
+S_{z_\alpha}\otimes S_{x_\alpha}\otimes (\widehat{V}_\alpha S_{y_\alpha})
-E^* S_{z_\alpha}\otimes S_{x_\alpha}\otimes S_{y_\alpha},
\end{eqnarray}
\begin{equation}
L_{\alpha M'M'\pm1} = \pm\widehat{\lambda}^{J,\pm M'}D_{z_\alpha}^{M'\pm}\otimes S_{x_\alpha}\otimes (Y_{\alpha r2}S_{y_\alpha}).
\end{equation}
Here we have introduced the ``one-dimensional'' matrices with elements
\begin{eqnarray}
(S_{z_\alpha})_{\zeta k} & = & S_{\alpha}^k(z_\alpha^\zeta),\quad
(S_{x_\alpha})_{\xi i} = S_{\alpha}^i(x_\alpha^\xi),\quad
(S_{y_\alpha})_{\eta j} = S_{\alpha}^j(y_\alpha^\eta), \nonumber \\
(D_{x_\alpha})_{\xi i} & = & \left(\frac{d^2}{dx_\alpha^2}S_\alpha^i\right)(x_\alpha^\xi),\quad
(D_{y_\alpha})_{\eta j} = \left(\frac{d^2}{dy_\alpha^2}S_\alpha^j\right)(y_\alpha^\eta), \nonumber \\
(D_{z_\alpha}^{M'})_{\zeta k} & = & \left( \bm{d}_{z_\alpha}^{M'}  S_\alpha^k\right)(z_\alpha^\zeta),\quad
(D_{z_\alpha}^{M'\pm})_{\zeta k} = \left( \bm{d}_{z_\alpha}^{M'\pm}  S_\alpha^k\right)(z_\alpha^\zeta), \nonumber \\
(X_{\alpha r2})_{\xi i} & = & \delta_{\xi i}\frac{1}{(x_\alpha^\xi)^2},\quad
(Y_{\alpha r2})_{\eta j} = \delta_{\eta j}\frac{1}{(y_\alpha^\eta)^2},\quad
(V_\alpha)_{\xi i} = \delta_{\xi i}V_\alpha(x_\alpha^\xi),\quad
(\widehat{V}_\alpha)_{\eta j} = \delta_{\eta j}\widehat{V}_\alpha(y_\alpha^\eta)
\end{eqnarray}
where we have used the notation
\begin{eqnarray}
\label{d-def}
\bm{d}_{z_\alpha}^{M'} & = & \left(1-(z_\alpha)^2\right) \frac{d^2}{dz_\alpha^2}
- 2(M'+1)z_\alpha \frac{d}{dz_\alpha} - M'(M'+1), \nonumber \\
\bm{d}_{z_\alpha}^{M'+} & = & -\left(1-(z_\alpha)^2\right)\frac{d}{dz_\alpha}+2(M'+1)z_\alpha,\quad
\bm{d}_{z_\alpha}^{M'-} = - \frac{d}{dz_\alpha}.
\end{eqnarray}
Let us find the diagonal representation
\begin{equation}
\label{z-dec}
(\overline{W}_{z_\alpha}^{M'},W_{z_\alpha}^{M'},\widetilde{\Lambda}_{z_\alpha}^{M'})=\mathfrak{D}(D_{z_\alpha}^{M'},S_{z_\alpha}),\quad
M' = M_0,\ldots,J.
\end{equation}
We introduce block diagonal matrices
\begin{eqnarray}
\overline{\mathcal{W}}_{z_\alpha} & = & \mathrm{diag}
\left\{
\overline{W}_{z_\alpha}^{M_0}\otimes I\otimes I, \overline{W}_{z_\alpha}^{M_0+1}\otimes I\otimes I,\ldots,\overline{W}_{z_\alpha}^{J}\otimes I\otimes I
\right\}, \nonumber \\
\mathcal{W}_{z_\alpha} & = & \mathrm{diag}
\left\{
W_{z_\alpha}^{M_0}\otimes I\otimes I, W_{z_\alpha}^{M_0+1}\otimes I\otimes I,\ldots,W_{z_\alpha}^{J}\otimes I\otimes I
\right\}.
\end{eqnarray}
Then the matrix $L_\alpha$ can be written in the form
\begin{equation}
L_\alpha =
\overline{\mathcal{W}}_{z_\alpha}^{\,-1}L_\alpha^{xy}\mathcal{W}_{z_\alpha}^{-1},
\end{equation}
where $L_\alpha^{xy}$ is a block tridiagonal matrix with blocks $L_{\alpha M'M'-1}^{xy}$, $L_{\alpha M'M'}^{xy}$ and $L_{\alpha M'M'+1}^{xy}$ in row $M' = M_0,\ldots,J$,
\begin{eqnarray}
\label{L-exact}
L_{\alpha M'M'}^{xy} &= & I\otimes\big(
-S_{x_\alpha}\otimes D_{y_\alpha}-D_{x_\alpha}\otimes S_{y_\alpha}
+\left( J(J+1)-2M'^2 \right) S_{x_\alpha}\otimes(Y_{\alpha r2}S_{y_\alpha}) \nonumber \\
 & + & (V_\alpha S_{x_\alpha})\otimes S_{y_\alpha}+S_{x_\alpha}\otimes(\widehat{V}_\alpha S_{y_\alpha})
-E^*S_{x_\alpha}\otimes S_{y_\alpha}
\big) \nonumber \\
 & - & \widetilde{\Lambda}_{z_\alpha}^{M'}\otimes
\big(
S_{x_\alpha}\otimes(Y_{\alpha r2}S_{y_\alpha}) + (X_{\alpha r2}S_{x_\alpha})\otimes S_{y_\alpha}
\big),
\end{eqnarray}
\begin{equation}
\label{L-pm-exact}
L_{\alpha M'M'\pm1}^{xy} = \pm \widehat{\lambda}^{J,\pm M'}
\left(\overline{W}_{z_\alpha}^{M'}D_{z_\alpha}^{M'\pm}W_{z_\alpha}^{M'\pm1}\right)\otimes S_{x_\alpha}\otimes (Y_{\alpha r2}S_{y_\alpha}).
\end{equation}
Now in order to build the preconditioner we introduce the block diagonal matrix $\mathcal{L}=\mathrm{diag}\{\mathcal{L}_1,\mathcal{L}_2,\mathcal{L}_3\}$.
This matrix is the approximation of the matrix $L$ of the operator at the left-hand side of 3D FM equations~(\ref{FM-fin}).
Its inverse $\mathcal{L}^{-1}$ is used as a preconditioner of the matrix $H-E^*S$ in the eigenvalue problem~(\ref{FM-lin}).
The block $\mathcal{L}_\alpha$ is given by
\begin{equation}
\mathcal{L}_\alpha \equiv
\overline{\mathcal{W}}_{z_\alpha}^{\,-1}\mathcal{L}_\alpha^{xy}\mathcal{W}_{z_\alpha}^{-1},
\end{equation}
where $\mathcal{L}_\alpha^{xy}$ has exactly the same block structure as the matrix $L_\alpha^{xy}$ with blocks $\mathcal{L}_{\alpha M'M'-1}^{xy}\approx L_{\alpha M'M'-1}^{xy}$, $\mathcal{L}_{\alpha M'M'}^{xy}\approx L_{\alpha M'M'}^{xy}$ and $\mathcal{L}_{\alpha M'M'+1}^{xy}\approx L_{\alpha M'M'+1}^{xy}$ in row $M' = M_0,\ldots,J$.
To obtain these blocks we note that under some assumptions on the basis of functions $S_\alpha^k(z)$ and the choice of collocation points $z_\alpha^\xi$ given in~\ref{app} the following approximate equalities hold
\begin{eqnarray}
\label{prop-eq}
\widetilde{\Lambda}_{z_\alpha}^{M'} \approx \Lambda_{z_\alpha}^{M'} & \equiv & \mathrm{diag}\left\{ -M'(M'+1), -(M'+1)(M'+2),\ldots,-(M'+n_{z_\alpha}-1)(M'+n_{z_\alpha}) \right\}, \nonumber \\
\overline{W}_{z_\alpha}^{M'}D_{z_\alpha}^{M'\pm}W_{z_\alpha}^{M'\pm1} & \approx & \Lambda_{z_\alpha}^{M'\pm},
\end{eqnarray}
where
\begin{equation}
\Lambda_{z_\alpha}^{M'+}=
\left(
\begin{array}{ccccc}
0 & & & & \\
-\lambda^{M'+1,M'} & 0 & & & \\
 & -\lambda^{M'+2,M'} & 0 & & \\
 & & & \ddots & \\
 & & & -\lambda^{M'+n_{z_\alpha}-1,M'} & 0
\end{array}
\right),
\end{equation}
\begin{equation}
\Lambda_{z_\alpha}^{M'-}=
\left(
\begin{array}{ccccc}
0 & \lambda^{M',-M'} & & & \\
 & 0 & \lambda^{M'+1,-M'} & & \\
 &  & \ddots & & \\
 & & & 0 & \lambda^{M'+n_{z_\alpha}-2,-M'} \\
 & & & & 0
\end{array}
\right).
\end{equation}
The justification of approximate equalities~(\ref{prop-eq}) is given in~\ref{app}.
Then the expressions for matrices $\mathcal{L}_{\alpha M'M'-1}^{xy}$, $\mathcal{L}_{\alpha M'M'}^{xy}$, $\mathcal{L}_{\alpha M'M'+1}^{xy}$ are obtained from~(\ref{L-exact}),~(\ref{L-pm-exact}) by making substitutions $\widetilde{\Lambda}_{z_\alpha}^{M'} \to \Lambda_{z_\alpha}^{M'}$, $\overline{W}_{z_\alpha}^{M'}D_{z_\alpha}^{M'\pm}W_{z_\alpha}^{M'\pm1} \to \Lambda_{z_\alpha}^{M'\pm}$. They are
\begin{eqnarray}
\mathcal{L}_{\alpha M'M'}^{xy} &= & I\otimes\big(
-S_{x_\alpha}\otimes D_{y_\alpha}-D_{x_\alpha}\otimes S_{y_\alpha}
+\left( J(J+1)-2M'^2 \right) S_{x_\alpha}\otimes(Y_{\alpha r2}S_{y_\alpha}) \nonumber \\
 & + & (V_\alpha S_{x_\alpha})\otimes S_{y_\alpha}+S_{x_\alpha}\otimes(\widehat{V}_\alpha S_{y_\alpha})
-E^*S_{x_\alpha}\otimes S_{y_\alpha}
\big) \nonumber \\
 & - & \Lambda_{z_\alpha}^{M'}\otimes
\big(
S_{x_\alpha}\otimes(Y_{\alpha r2}S_{y_\alpha}) + (X_{\alpha r2}S_{x_\alpha})\otimes S_{y_\alpha}
\big),
\end{eqnarray}
\begin{equation}
\mathcal{L}_{\alpha M'M'\pm1}^{xy} = \pm \widehat{\lambda}^{J,\pm M'}
\Lambda_{z_\alpha}^{M'\pm}\otimes S_{x_\alpha}\otimes (Y_{\alpha r2}S_{y_\alpha}).
\end{equation}
In the remaining part of this subsection we deal with the approximate matrix $\mathcal{L}$ and show how to invert it efficiently.

We proceed further by noting that each diagonal block $\mathcal{L}_{\alpha M'M'}^{xy}$ of the matrix $\mathcal{L}_{\alpha}^{xy}$ has a block diagonal structure with ``two-dimensional'' blocks $\mathcal{M}_{\alpha\ell}^{M'}$ of the form
\begin{eqnarray}
\mathcal{M}_{\alpha \ell}^{M'} &= &
S_{x_\alpha}\otimes\big[
-D_{y_\alpha}
+\left( J(J+1)-2M'^2 \right) (Y_{\alpha r2}S_{y_\alpha})  \nonumber \\
 & + & \ell(\ell+1)(Y_{\alpha r2}S_{y_\alpha}) +  (\widehat{V}_\alpha S_{y_\alpha})
-E^* S_{y_\alpha} \big] \nonumber \\
 & + & \big[-D_{x_\alpha}  + \ell(\ell+1) 
 (X_{\alpha r2}S_{x_\alpha}) + (V_\alpha S_{x_\alpha})\big]\otimes S_{y_\alpha},\quad
\ell = M',\ldots,M'+n_{z_\alpha}-1.
\end{eqnarray}
Likewise, the off-diagonal blocks $\mathcal{L}_{\alpha M'M'\pm1}^{xy}$ are block matrices with nontrivial blocks $\mathcal{M}_{\alpha\ell}^{M'\pm}$ on subdiagonal of $\mathcal{L}_{\alpha M'M'+1}^{xy}$ and superdiagonal of $\mathcal{L}_{\alpha M'M'-1}^{xy}$.
They have the form
\begin{eqnarray}
\mathcal{M}_{\alpha\ell}^{M'+} & = & \widehat{\lambda}^{J,M'}
\left(
\Lambda_{z_\alpha}^{M'+}
\right)_{\ell-M'+1,\ell-M'}
S_{x_\alpha}\otimes(Y_{\alpha r2}S_{y_\alpha}),\quad \ell=M'+1,\ldots,M'+n_{z_\alpha}-1, \nonumber \\
\mathcal{M}_{\alpha\ell}^{M'-} & = & -\widehat{\lambda}^{J,-M'}
\left(
\Lambda_{z_\alpha}^{M'-}
\right)_{\ell-M'+1,\ell-M'+2}
S_{x_\alpha}\otimes(Y_{\alpha r2}S_{y_\alpha}),\quad \ell=M',\ldots,M'+n_{z_\alpha}-2.
\end{eqnarray}
Thus the matrix $\mathcal{L}_\alpha^{xy}$ has the form shown in Fig.~\ref{l-alpha} with blocks with indices ${}_{\ell}^{M'}$ and ${}_{\ell}^{M'\pm}$ denoting matrices $\mathcal{M}_{\alpha\ell}^{M'}$, $\mathcal{M}_{\alpha\ell}^{M'\pm}$.
\begin{figure}[!h]
\[
\begin{BMAT}(e)[0pt]{cccc}{cccc}
\begin{BMAT}(@,41pt,41pt){cccc}{cccc}
{}_{ M_0}^{M_0} &  &  &  \\
 & {}_{ M_0+1}^{M_0} &  &  \\
 &  & \ddots &  \\
 &  &  & {}_{ M_0+n_{z_\alpha}-1}^{M_0}
\addpath{(1,3,3)uldr}
\addpath{(2,2,3)uldr}
\addpath{(4,0,3)uldr}
\end{BMAT} & 
\begin{BMAT}(@,41pt,41pt){cccc}{cccc}
 &  &  &  \\
{}_{ M_0+1}^{M_0+} &  &  &  \\
 & \ddots &  &  \\
 &  & {}_{ M_0+n_{z_\alpha}-1}^{M_0+} &
\addpath{(1,2,3)uldr}
\addpath{(3,0,3)uldr}
\end{BMAT} &  &  \\
\begin{BMAT}(@,41pt,41pt){cccc}{cccc}
 & {}_{ M_0+1}^{(M_0+1)-} &  &  \\
 &  & \ddots &  \\
 &  &  & {}_{ M_0+n_{z_\alpha}-1}^{(M_0+1)-}  \\
 &  &  &
\addpath{(2,3,3)uldr}
\addpath{(4,1,3)uldr}
\end{BMAT} &
\begin{BMAT}(@,41pt,41pt){cccc}{cccc}
{}_{ M_0+1}^{M_0+1} &  &  &  \\
 & {}_{ M_0+2}^{M_0+1} &  &  \\
 &  & \ddots &  \\
 &  &  & {}_{ M_0+n_{z_\alpha}}^{M_0+1}
\addpath{(1,3,3)uldr}
\addpath{(2,2,3)uldr}
\addpath{(4,0,3)uldr}
\end{BMAT} & 
\begin{BMAT}(@,41pt,41pt){cccc}{cccc}
 &  &  &  \\
{}_{ M_0+2}^{(M_0+1)+} &  &  &  \\
 & \ddots &  &  \\
 &  & {}_{ M_0+n_{z_\alpha}}^{(M_0+1)+} &
\addpath{(1,2,3)uldr}
\addpath{(3,0,3)uldr}
\end{BMAT} &  \\
\begin{BMAT}(@,41pt,41pt){cccc}{cccc}
 &  &  &  \\
\cdots & \cdots & \cdots  & \cdots \\
\cdots & \cdots & \cdots & \cdots \\
 &  &  & 
\end{BMAT} &
\begin{BMAT}(@,41pt,41pt){cccc}{cccc}
 &  &  &  \\
\cdots & \cdots & \cdots  & \cdots \\
\cdots & \cdots & \cdots & \cdots \\
 &  &  & 
\end{BMAT} & 
\begin{BMAT}(@,41pt,41pt){cccc}{cccc}
 &  &  &  \\
\cdots & \cdots & \cdots  & \cdots \\
\cdots & \cdots & \cdots & \cdots \\
 &  &  & 
\end{BMAT} & \\
 & 
\begin{BMAT}(@,41pt,41pt){cccc}{cccc}
 & {}_{ J}^{J-} &  &  \\
 &  & \ddots &  \\
 &  &  & {}_{ J+n_{z_\alpha}-2}^{J-}  \\
 &  &  &
\addpath{(2,3,3)uldr}
\addpath{(4,1,3)uldr}
\end{BMAT} & 
\begin{BMAT}(@,41pt,41pt){cccc}{cccc}
{}_{ J}^{J} &  &  &  \\
 & {}_{ J+1}^{J} &  &  \\
 &  & \ddots &  \\
 &  &  & {}_{ J+n_{z_\alpha}-1}^{J}
\addpath{(1,3,3)uldr}
\addpath{(2,2,3)uldr}
\addpath{(4,0,3)uldr}
\end{BMAT} &
\begin{BMAT}(@,41pt,41pt){cccc}{cccc}
 &  &  &  \\
 &  &  &  \\
 &  &  &  \\
 &  &  & 
\end{BMAT}
\addpath{(1,3,3)uldr}
\addpath{(2,3,3)uldr}
\addpath{(1,2,3)uldr}
\addpath{(2,2,3)uldr}
\addpath{(3,2,3)uldr}
\addpath{(2,0,3)uldr}
\addpath{(3,0,3)uldr}
\end{BMAT}
\]
\caption{Structure of the matrices $\mathcal{L}_\alpha^{xy}$, $\mathcal{L}_\alpha^{y}$ and $\mathcal{L}_\alpha^0$.}
\label{l-alpha}
\end{figure}
Now let
\begin{equation}
(\overline{W}_{x_\alpha}^\ell,W_{x_\alpha}^\ell,\Lambda_{x_\alpha}^\ell) = 
\mathfrak{D}(-D_{x_\alpha}+\ell(\ell+1)X_{\alpha r2}S_{x_\alpha}+V_\alpha S_{x_\alpha}, S_{x_\alpha}),\quad
\ell=M_0,\ldots,J+n_{z_\alpha}-1.
\end{equation}
Introduce new block diagonal matrices
\begin{eqnarray}
\overline{\mathcal{W}}_{x_\alpha} & = &
\mathrm{diag}
\left\{
\overline{\mathcal{W}}_{x_\alpha}^{M_0},\ldots,\overline{\mathcal{W}}_{x_\alpha}^{M_0+n_{z_\alpha}-1},
\overline{\mathcal{W}}_{x_\alpha}^{M_0+1},\ldots,\overline{\mathcal{W}}_{x_\alpha}^{M_0+n_{z_\alpha}},
\ldots\ldots\ldots,\overline{\mathcal{W}}_{x_\alpha}^{J},\ldots,\overline{\mathcal{W}}_{x_\alpha}^{J+n_{z_\alpha}-1}
\right\},
\nonumber \\
\mathcal{W}_{x_\alpha} & = &
\mathrm{diag}
\left\{
\mathcal{W}_{x_\alpha}^{M_0},\ldots,\mathcal{W}_{x_\alpha}^{M_0+n_{z_\alpha}-1},
\mathcal{W}_{x_\alpha}^{M_0+1},\ldots,\mathcal{W}_{x_\alpha}^{M_0+n_{z_\alpha}},
\ldots\ldots\ldots,\mathcal{W}_{x_\alpha}^{J},\ldots,\mathcal{W}_{x_\alpha}^{J+n_{z_\alpha}-1}
\right\},
\end{eqnarray}
where $\overline{\mathcal{W}}_{x_\alpha}^{\ell} = \overline{W}_{x_\alpha}^\ell\otimes I$ and $\mathcal{W}_{x_\alpha}^{\ell} = W_{x_\alpha}^\ell\otimes I$. 
Then one can check that
\begin{equation}
\mathcal{L}_\alpha^{xy} =
\overline{\mathcal{W}}_{x_\alpha}^{\,-1}\mathcal{L}_\alpha^y\mathcal{W}_{x_\alpha}^{-1},
\end{equation}
where the new matrix $\mathcal{L}_\alpha^y$ has exactly the same form as $\mathcal{L}_\alpha^{xy}$ shown in Fig.~\ref{l-alpha} with blocks $\mathcal{N}_{\alpha\ell}^{M'}$, $\mathcal{N}_{\alpha\ell}^{M'\pm}$.
These new blocks are
\begin{eqnarray}
\mathcal{N}_{\alpha\ell}^{M'} & = & I\otimes
\left(
-D_{y_\alpha}+\left( J(J+1)-2M'^2+\ell(\ell+1) \right)(Y_{\alpha r2}S_{y_\alpha})+\widehat{V}_\alpha S_{y_\alpha}-E^*S_{y_\alpha}
\right)
+\Lambda_{x_\alpha}^\ell\otimes S_{y_\alpha},
\nonumber \\
\mathcal{N}_{\alpha\ell}^{M'+} & = & \widehat{\lambda}^{J,M'}
\left(
\Lambda_{z_\alpha}^{M'+}
\right)_{\ell-M'+1,\ell-M'}
I\otimes (Y_{\alpha r2}S_{y_\alpha}),\quad \ell=M'+1,\ldots,M'+n_{z_\alpha}-1, \nonumber \\
\mathcal{N}_{\alpha\ell}^{M'-} & = & -\widehat{\lambda}^{J,-M'}
\left(
\Lambda_{z_\alpha}^{M'-}
\right)_{\ell-M'+1,\ell-M'+2}
I\otimes(Y_{\alpha r2}S_{y_\alpha}),\quad \ell=M',\ldots,M'+n_{z_\alpha}-2.
\end{eqnarray}
They are in turn block diagonal
\begin{eqnarray}
\mathcal{N}_{\alpha\ell}^{M'} & = & \mathrm{diag}\{
\mathcal{N}_{\alpha\ell1}^{M'},\mathcal{N}_{\alpha\ell2}^{M'},\ldots,\mathcal{N}_{\alpha\ell n_{x_\alpha}}^{M'}
\}
\nonumber \\
\mathcal{N}_{\alpha\ell}^{M'\pm} & = & \mathrm{diag}\{
\mathcal{N}_{\alpha\ell0}^{M'\pm},\mathcal{N}_{\alpha\ell0}^{M'\pm},\ldots,\mathcal{N}_{\alpha\ell 0}^{M'\pm}
\},
\end{eqnarray}
where the ``one-dimensional'' matrices have the form
\begin{eqnarray}
\mathcal{N}_{\alpha\ell i_{x_\alpha}}^{M'} & = &
-D_{y_\alpha}+\widehat{V}_\alpha S_{y_\alpha}-E^* S_{y_\alpha}+
\left(
\Lambda_{x_\alpha}^\ell
\right)_{i_{x_\alpha},i_{x_\alpha}} S_{y_\alpha}\nonumber \\
 & + & \left( J(J+1)-2M'^2+\ell(\ell+1) \right)(Y_{\alpha r2} S_{y_\alpha}),\quad
\ell=M',\ldots,M'+n_{z_\alpha}-1,\quad i_{x_\alpha}=1,\ldots,n_{x_\alpha},
\nonumber \\
\mathcal{N}_{\alpha\ell 0}^{M'+} & = & \widehat{\lambda}^{J,M'}
\left(
\Lambda_{z_\alpha}^{M'+}
\right)_{\ell-M'+1,\ell-M'}
(Y_{\alpha r2}S_{y_\alpha}),\quad \ell=M'+1,\ldots,M'+n_{z_\alpha}-1, \nonumber \\
\mathcal{N}_{\alpha\ell 0}^{M'-} & = & -\widehat{\lambda}^{J,-M'}
\left(
\Lambda_{z_\alpha}^{M'-}
\right)_{\ell-M'+1,\ell-M'+2}
(Y_{\alpha r2}S_{y_\alpha}),\quad \ell=M',\ldots,M'+n_{z_\alpha}-2.
\end{eqnarray}
Now we make the last step of diagonalization process.
Introduce
\begin{eqnarray}
\label{ops}
(\overline{W}_{y_\alpha}^{\ell i_{x_\alpha}},W_{y_\alpha}^{\ell i_{x_\alpha}},\Lambda_{y_\alpha}^{\ell i_{x_\alpha}})
& = & \mathcal{D}
\left(
-D_{y_\alpha}+\widehat{V}_\alpha S_{y_\alpha}-E^*S_{y_\alpha}+\left( \Lambda_{x_\alpha}^\ell \right)_{i_{x_\alpha}i_{x_\alpha}}S_{y_\alpha},
Y_{\alpha r2}S_{y_\alpha}
\right),\nonumber \\
 & & \ell=M_0,\ldots,J+n_{z_\alpha}-1,\quad
i_{x_\alpha}=1,\ldots,n_{x_\alpha}.
\end{eqnarray}
Then using the matrices
\begin{eqnarray}
\overline{\mathcal{W}}_{y_\alpha} & = &
\mathrm{diag}\{
\overline{\mathcal{W}}_{y_\alpha}^{M_0},\ldots,\overline{\mathcal{W}}_{y_\alpha}^{M_0+n_{z_\alpha}-1},
\overline{\mathcal{W}}_{y_\alpha}^{M_0+1},\ldots,\overline{\mathcal{W}}_{y_\alpha}^{M_0+n_{z_\alpha}},
\ldots\ldots\ldots,
\overline{\mathcal{W}}_{y_\alpha}^{J},\ldots,\overline{\mathcal{W}}_{y_\alpha}^{J+n_{z_\alpha}-1}
\}, \nonumber \\
\mathcal{W}_{y_\alpha} & = &
\mathrm{diag}\{
\mathcal{W}_{y_\alpha}^{M_0},\ldots,\mathcal{W}_{y_\alpha}^{M_0+n_{z_\alpha}-1},
\mathcal{W}_{y_\alpha}^{M_0+1},\ldots,\mathcal{W}_{y_\alpha}^{M_0+n_{z_\alpha}},
\ldots\ldots\ldots,
\mathcal{W}_{y_\alpha}^{J},\ldots,\mathcal{W}_{y_\alpha}^{J+n_{z_\alpha}-1}
\},
\end{eqnarray}
where $\overline{\mathcal{W}}_{y_\alpha}^{\ell}$ and $\mathcal{W}_{y_\alpha}^{\ell}$ are in turn block diagonal matrices of the form
\begin{eqnarray}
\overline{\mathcal{W}}_{y_\alpha}^{\ell} & = &
\mathrm{diag}
\left\{
\overline{W}_{y_\alpha}^{\ell 1},\ldots,\overline{W}_{y_\alpha}^{\ell n_{x_\alpha}}
\right\},
\nonumber \\
\mathcal{W}_{y_\alpha}^{\ell} & = &
\mathrm{diag}
\left\{
W_{y_\alpha}^{\ell 1},\ldots,W_{y_\alpha}^{\ell n_{x_\alpha}}
\right\},
\end{eqnarray}
we represent
\begin{equation}
\mathcal{L}_\alpha^y = \overline{\mathcal{W}}_{y_\alpha}^{\,-1}
\mathcal{L}_\alpha^0
\mathcal{W}_{y_\alpha}^{-1}.
\end{equation}
Here the matrix $\mathcal{L}_\alpha^0$ has exactly the same form as $\mathcal{L}_\alpha^{xy}$ and $\mathcal{L}_\alpha^y$ shown in Fig.~\ref{l-alpha} with blocks $\mathcal{P}_{\alpha\ell}^{M'}$, $\mathcal{P}_{\alpha\ell}^{M'\pm}$.
These new blocks are diagonal matrices with elements
\begin{eqnarray}
\left(\mathcal{P}_{\alpha\ell}^{M'}\right)_{(i_{x_\alpha}-1)n_{y_\alpha}+i_{y_\alpha},(i_{x_\alpha}-1)n_{y_\alpha}+i_{y_\alpha}} & = &
\left( \Lambda_{y_\alpha}^{\ell,i_{x_\alpha}} \right)_{i_{y_\alpha},i_{y_\alpha}}
+J(J+1)-2M'^2+\ell(\ell+1),\nonumber \\
\left(\mathcal{P}_{\alpha\ell}^{M'+}\right)_{(i_{x_\alpha}-1)n_{y_\alpha}+i_{y_\alpha},(i_{x_\alpha}-1)n_{y_\alpha}+i_{y_\alpha}} & = &
\widehat{\lambda}^{J,M'}
\left(
\Lambda_{z_\alpha}^{M'+}
\right)_{\ell-M'+1,\ell-M'},
\nonumber \\
\left(\mathcal{P}_{\alpha\ell}^{M'-}\right)_{(i_{x_\alpha}-1)n_{y_\alpha}+i_{y_\alpha},(i_{x_\alpha}-1)n_{y_\alpha}+i_{y_\alpha}} & = &
-\widehat{\lambda}^{J,-M'}
\left(
\Lambda_{z_\alpha}^{M'-}
\right)_{\ell-M'+1,\ell-M'+2},
\nonumber \\
 & & i_{x_\alpha}=1,\ldots,n_{x_\alpha},\quad i_{y_\alpha}=1,\ldots,n_{y_\alpha}.
\end{eqnarray}
In other words the matrix $\mathcal{L}_\alpha^0$ is a band matrix with three nonzero diagonals.
It is easy to show that there exists a permutation matrix $P$ such that
\begin{equation}
\mathcal{L}_\alpha^0 = P\left( P\mathcal{L}_\alpha^0P \right)P = P\widetilde{\mathcal{L}}_\alpha^0P,
\end{equation}
where $\widetilde{\mathcal{L}}_\alpha^0$ is block diagonal with each block being a three-diagonal matrix.
Every block comprises rows of $\mathcal{L}_\alpha^0$ with numbers forming sets of numbers with equal remainders of the division by $(n_{z_\alpha}-1)n_{x_\alpha}n_{y_\alpha}$.
Thus the matrix $\mathcal{L}_\alpha^0$ can be inverted fast.

Summarizing, we have obtained
\begin{equation}
\mathcal{L}_\alpha =
\overline{\mathcal{W}}_{z_\alpha}^{\,-1}\overline{\mathcal{W}}_{x_\alpha}^{\,-1}\overline{\mathcal{W}}_{y_\alpha}^{\,-1}
P\widetilde{\mathcal{L}}_\alpha^0P
\mathcal{W}_{y_\alpha}^{-1}\mathcal{W}_{x_\alpha}^{-1}\mathcal{W}_{z_\alpha}^{-1}
\end{equation}
and the approximate preconditioner can be presented in the form
\begin{equation}
\label{prec-1}
\mathcal{L}^{-1}=\mathrm{diag}
\{
\mathcal{L}_1^{-1},\mathcal{L}_2^{-1},\mathcal{L}_3^{-1}
\},
\end{equation}
with
\begin{equation}
\label{prec-2}
\mathcal{L}_\alpha^{-1} =
\mathcal{W}_{z_\alpha}\mathcal{W}_{x_\alpha}\mathcal{W}_{y_\alpha}
P\left(\widetilde{\mathcal{L}}_\alpha^0\right)^{-1}P
\overline{\mathcal{W}}_{y_\alpha}\overline{\mathcal{W}}_{x_\alpha}\overline{\mathcal{W}}_{z_\alpha}.
\end{equation}
The operations~(\ref{ops}) are the most computer time and memory consuming operations in calculating and storing the preconditioner.
It is needed to store and calculate the total amount of $\sum_\alpha (n_M+n_{z_\alpha}-1)n_{x_\alpha}$ pairs $(\overline{W}_{y_\alpha}^{\ell,i_{x_\alpha}},W_{y_\alpha}^{\ell,i_{x_\alpha}})$ of matrices of general form.
Thus the estimated memory requirements are $2\sum_\alpha (n_{z_\alpha}+n_M)n_{x_\alpha}n_{y_\alpha}^2$ numbers and the computational cost scales as $O(\sum_\alpha (n_{z_\alpha}+n_M)n_{x_\alpha}n_{y_\alpha}^3)$.
Finally, each matrix-vector product with the preconditioner $\mathcal{L}^{-1}$ requires $O(n_M\sum_\alpha n_{x_\alpha}n_{z_\alpha}n_{y_\alpha}(n_{x_\alpha}+n_{z_\alpha}+n_{y_\alpha}))$ operations.

\subsection{Matrix storage}
\label{storage}

In the last part of this section we discuss the storage of the full matrix $H$ which is the most memory consuming part of the scheme.
In this discussion it is more convenient to refer to the 3D FM equations written in the form~(\ref{FM-pre}).
Let $H=\widetilde{L}+\widetilde{R}$ with $\widetilde{L}$ and $\widetilde{R}$ being the discretized versions of the operators at the left- and right-hand sides of 3D FM equations~(\ref{FM-pre}).
The matrices $\widetilde{L}$ and $\widetilde{R}$ contain diagonal and off-diagonal with respect to component numbers $\alpha$ blocks of the full matrix $H$ and thus can be stored independently.
Now one obvious way is to keep those matrices in memory as sparse matrices using one of the well known storage schemes like the CSR format~\cite{Saad03}.
The estimated storage size and computational cost of matrix-vector product are $(3n_M-2)r^3\sum_\alpha n_{x_\alpha}n_{y_\alpha}n_{z_\alpha}$ numbers and multiplications for $\widetilde{L}$ and $2n_M^2r^3\sum_\alpha n_{x_\alpha}n_{y_\alpha}n_{z_\alpha}$ for $\widetilde{R}$.

These requirements can be drastically lowered in the case of using the same basis set for expanding components $\widehat{\psi}_{\alpha MM'}^{J\tau}$ with different $M'$ and identical sets of collocation points for equations~(\ref{FM-fin}) with different $M'$.
One possible storage scheme that we use in the calculations is presented here.
The blocks $\widetilde{L}_\alpha$ of the matrix $\widetilde{L}$ are stored in the form
\begin{eqnarray}
\label{l-form}
\widetilde{L}_\alpha & = &
I_M\otimes D_{x_\alpha y_\alpha z_\alpha} + MS_{z_\alpha}\otimes SYS_{x_\alpha y_\alpha}+MD_{z_\alpha}\otimes XSYS_{x_\alpha y_\alpha}.
\end{eqnarray}
Here the matrix $I_M$ is the $n_M\times n_M$ identity matrix in the space of momenta numbers $M'$.
The matrices that are stored in memory as sparse matrices in CSR format are: $D_{x_\alpha y_\alpha z_\alpha}$ is the discretized version of the three-dimensional operator
\begin{equation}
-\frac{\partial^2}{\partial y_\alpha^2}-\frac{\partial^2}{\partial x_\alpha^2}
-\left(\frac{1}{y_\alpha^2}+\frac{1}{x_\alpha^2}\right)(1-z_\alpha^2)\frac{\partial^2}{\partial z_\alpha^2}
+V_\alpha(x_\alpha)+\sum_{\beta\ne\alpha}V_\beta^{(\mathrm{l})}(x_\beta(X_\alpha),y_\beta(X_\alpha))-E
\end{equation}
and other matrices have the form
\begin{eqnarray}
SYS_{x_\alpha y_\alpha} & = & S_{x_\alpha} \otimes (Y_{\alpha r2}S_{y_\alpha}), \nonumber \\
XSYS_{x_\alpha y_\alpha} & = & (X_{\alpha r2}S_{x_\alpha}) \otimes S_{y_\alpha}+S_{x_\alpha} \otimes (Y_{\alpha r2}S_{y_\alpha}), \nonumber \\
MS_{z_\alpha} & = & JM \otimes S_{z_\alpha} + \Lambda_M^-\otimes D_{1z_\alpha} - \Lambda_M^+\otimes \left( (I-Z_\alpha^2)D_{1z_\alpha}\right)+2\left((MM+I_M)\Lambda_M^+\right) \otimes (Z_\alpha S_{z_\alpha}),\nonumber \\
MD_{z_\alpha} & = & 2(MM+I_M)\otimes (Z_\alpha D_{1z_\alpha}) + \left(MM(MM+I_M)\right) \otimes S_{z_\alpha}.
\end{eqnarray}
Here we have additionally denoted by $D_{1z_\alpha}$ and $Z_\alpha$ the matrices with elements
\begin{eqnarray}
\left(
D_{1z_\alpha}
\right)_{\zeta k} & = & 
\left(\frac{d}{dz_\alpha}S_\alpha^k\right)(z_\alpha^\zeta),\quad
\left(
Z_\alpha
\right)_{\zeta k} = 
\delta_{\zeta k} z_\alpha^\zeta
\end{eqnarray}
and the matrices in the space of momenta numbers are
\begin{eqnarray}
JM & = & \mathrm{diag}\left\{ J(J+1)-2M_0^2, J(J+1)-2(M_0+1)^2, \ldots, J(J+1)-2J^2 \right\},\nonumber \\
MM & = & \mathrm{diag}\left\{ M_0,M_0+1,\ldots,J\right\},\nonumber \\
\Lambda_M^-& = &
\left(
\begin{array}{ccccc}
0 & & & & \\
\widehat{\lambda}^{J,-(M_0+1)} & 0 & & & \\
 & \widehat{\lambda}^{J,-(M_0+2)} & 0 & & \\
 & & & \ddots & \\
 & & & \widehat{\lambda}^{J,-J} & 0
\end{array}
\right), \nonumber \\
\Lambda_M^+& = &
\left(
\begin{array}{ccccc}
0 & \widehat{\lambda}^{J,M_0} & & & \\
 & 0 & \widehat{\lambda}^{J,M_0+1} & & \\
 &  & \ddots & & \\
 & & & 0 & \widehat{\lambda}^{J,J-1} \\
 & & & & 0
\end{array}
\right).
\end{eqnarray}
If blocks of the matrix $\widetilde{L}$ are stored in the form~(\ref{l-form}), then it requires a storage of $O(r^3\sum_\alpha n_{x_\alpha} n_{y_\alpha} n_{z_\alpha} + 4n_M r \sum_\alpha n_{z_\alpha} + 2r^2\sum_\alpha n_{x_\alpha}n_{y_\alpha})$ numbers and it takes $O(n_M(r^3 + 2r^2+4r)\sum_\alpha n_{x_\alpha} n_{y_\alpha} n_{z_\alpha})$ multiplication operations for matrix-vector product.

The matrix $\widetilde{R}$ is a block matrix with blocks $\widetilde{R}_{\alpha\beta}$. If no permutational symmetry is taken into account, it has trivial diagonal blocks $\widetilde{R}_{\alpha\beta}=0,\alpha=\beta$.
The nonzero blocks $\widetilde{R}_{\alpha\beta}$ can be stored in the form
\begin{equation}
\label{r-form}
\widetilde{R}_{\alpha\beta} = F_{\alpha\beta} \widetilde{S}_{\alpha\beta},
\end{equation}
with
\begin{equation}
\label{s-rot}
\widetilde{S}_{\alpha\beta} = I_M \otimes S_{\alpha\beta},
\end{equation}
where the ``three-dimensional'' matrix $S_{\alpha\beta}$ that is stored in memory as CSR sparse matrix has elements
\begin{equation}
\left(
S_{\alpha\beta}
\right)_{\zeta\xi\eta,kij}=
S_\beta^k(z_\beta(x_\alpha^\xi,y_\alpha^\eta,z_\alpha^\zeta))
S_\beta^i(x_\beta(x_\alpha^\xi,y_\alpha^\eta,z_\alpha^\zeta))
S_\beta^j(y_\beta(x_\alpha^\xi,y_\alpha^\eta,z_\alpha^\zeta)).
\end{equation}
Note that if different basis sets are used for expanding components $\widehat{\psi}_{\alpha MM'}^{J\tau}$, the decomposition~(\ref{r-form}) still makes sense, but $\widetilde{S}_{\alpha\beta}$ cannot be presented in the form~(\ref{s-rot}). Instead, it is then block diagonal with $n_M$ different blocks. 
Each stored in memory matrix $F_{\alpha\beta}$ is a block matrix with blocks $F_{\alpha\beta}^{M'M''}$, $M',M''=M_0,\ldots,J$. Each block is a ``three-dimensional'' diagonal matrix with elements
\begin{eqnarray}
\left(
F_{\alpha\beta}^{M'M''}
\right)_{\zeta\xi\eta,kij} & = & 
-\delta_{\zeta k}\delta_{\xi i}\delta_{\eta j}
\frac{x_\alpha^\xi y_\alpha^\eta}{x_\beta(x_\alpha^\xi,y_\alpha^\eta,z_\alpha^\zeta)y_\beta(x_\alpha^\xi,y_\alpha^\eta,z_\alpha^\zeta)}
\frac{\left( 1-(z_\beta(x_\alpha^\xi,y_\alpha^\eta,z_\alpha^\zeta))^2 \right)^{M''/2}}{\left( 1-(z_\alpha)^2 \right)^{M'/2}}
V_\alpha^{(\mathrm{s})}(x_\alpha^\xi,y_\alpha^\eta)\nonumber \\
 & \times & (-1)^{M''-M'}\frac{2}{\sqrt{2+2\delta_{M''0}}}
F_{M''M'}^{J\tau}(0,w_{\beta\alpha},0).
\end{eqnarray}
The matrix $F_{\alpha\beta}$ is stored in memory as a set of diagonal matrices.
Storing blocks of matrix $\widetilde{R}$ in the form~(\ref{r-form}) requires storage of $O\left(\left(n_M^2 + r^3 \right) \sum_\alpha n_{x_\alpha}n_{y_\alpha}n_{z_\alpha}\right)$ numbers and the same number of multiplication operations for matrix-vector product.
This is the drastical progress as compared with storing $\widetilde{R}$ as sparse matrix of general form.
However, simple analysis shows that using the forms~(\ref{l-form}) and~(\ref{r-form}) gives advantage compared with the full matrix CSR storage only if $n_M$ is greater than one.

\section{Examples}

In our realization of the described method the basis functions are in the space of quintic Hermite splines $S_5^3$ (splines of degree 5 with 2 continuous derivatives).
Each basis function is local and nonzero only on two adjoining intervals of the grid.
There are three functions associated with each grid node and thus for this choice of basis the defined in previous section parameter $r=6$.
The spline nodes on the solution interval are chosen by mapping the equidistant grid with some function.
The mapping function~\cite{Korn19}
\begin{equation}
\sin(z_\alpha(i) \pi/2),
\end{equation}
where $z_\alpha(i)$ are points of the equidistant grid on interval $[-1,1]$,
is used for splines in coordinate $z_\alpha$ in all calculations except those with high values of total orbital momentum $J$ presented in Table~\ref{He-10}. In this case the more suited equidistant nodes have been used.
For splines in coordinates $x_\alpha$ and $y_\alpha$ the mapping functions are chosen depending on the physical problem being solved.
We use collocation at Gaussian points~\cite{Boor73, Bial01}.
The largest modulus eigenvalues of the eigenvalue problem~(\ref{FM-lin}) are obtained by Implicitly Restarted Arnoldi Method (IRAM)~\cite{Soren92, Saad11}.
We use the algorithm implemented in ARPACK~\cite{Arpack98} realization of IRAM.
In the process of IRAM iterations the matrix $(H-E^*S)$ of~(\ref{FM-lin}) is being inverted by preconditioned GMRES algorithm~\cite{Saad03}.
Complex arithmetic is used in the calculations.
The realization is written in C++ programming language with the use of Intel Parallel Studio XE 2019 Update 4 software package.
Parallelization of the program is achieved by using functions from Intel MKL~\cite{MKL} mathematical library.
All calculations are done on a 6 core machine with Intel Xeon X5675 processor with 32 Gbytes RAM.
In the presentation of results we adapt the notation
\begin{equation}
(\alpha:n_{x_\alpha},n_{y_\alpha},n_{z_\alpha};\ \beta:n_{x_\beta},n_{y_\beta},n_{z_\beta}),\quad (\alpha:R_{x_\alpha},R_{y_\alpha};\ \beta:R_{x_\beta},R_{y_\beta})
\end{equation}
for the basis sizes used to represent the partial components $\widehat{\psi}_{\alpha MM'}^{J\tau}$ and lengths of intervals they are defined on.

We used two physical systems for the test calculations of bound state energies: the Helium atom and the molecular Helium trimer.
All calculations for the Helium atom were conducted in atomic units.
Since we have taken the results of~\cite{aznab15} as benchmark results for atomic Helium energies, infinite ($10^{15}$ a.u. in practice) Helium core mass was used in the calculations as in~\cite{aznab15}.
To generate nodes of spline in variable $x_\alpha$ ($y_\alpha$) on interval $[0, R_{x_\alpha}]$ ($[0, R_{y_\alpha}]$) we employed the mapping~\cite{Kvits93}
\begin{equation}
\label{map-pow}
4\left( \left(1+\frac{R_{x_\alpha}(R_{y_\alpha})}{4}\right)^{u(i)}-1 \right),
\end{equation}
with $u(i)=x_\alpha(i)/R_{x_\alpha}(y_\alpha(i)/R_{y_\alpha})$ and equidistant points $x_\alpha(i)$ ($y_\alpha(i)$).
Splitting of the Coulomb potentials were done with the Merkuriev cut-off function of the form~(\ref{Mcutoff0}) with $\nu_\alpha=2.01$, $x_{0\alpha}=1$ a.u. and $y_{0\alpha}=+\infty$.
The potentials $\widehat{V}_\alpha$ in equations~(\ref{FM-fin}) are chosen to be
\begin{equation}
\widehat{V}_\alpha(y_\alpha) = \frac{Z_\alpha(Z_\beta+Z_\gamma)}{\sqrt{2\mu_{\alpha(\beta\gamma)}}y_\alpha}
\left(
1-e^{\left(-\frac{y_\alpha}{0.05R_{y_\alpha}}\right)^2}
\right).
\end{equation}
The permutational symmetry with respect to electrons have been used in the calculations by reducing the number of equations~(\ref{FM-fin}) to $2n_M$ for a given orbital momentum $J$.

The Helium trimer is a weakly-bound molecular system that has only two bound states with $L=0$ with extremely small energy values and spatially extended wave functions.
Calculations of these energy levels is a well-known computational challenge.
We compare our test calculation results with benchmark calculations of~\cite{Roudn12}.
The interatomic interactions are described by the TTY potential and atomic Helium mass is taken to be $M_{\mathrm{He}}=7296.2994$ a.u. To normalize the values of coordinates $x_\alpha$ and $y_\alpha$ in the calculations we used $1/M_{\mathrm{He}}$ a.u. as energy units and $\sqrt{M_{\mathrm{He}}}$ a.u. as length units.
For $x_\alpha$ spline nodes we used the mapping~(\ref{map-pow}) and for $y_\alpha$
\begin{equation}
\left(  \left( \left( R_{y_\alpha}+1 \right)^{\frac13} -1 \right)u(i) + 1 \right)^3-1
\end{equation}
with $u(i)=y_\alpha(i)/R_{y_\alpha}$ and equidistant points $y_\alpha(i)$.
The permutational symmetry with respect to all three particles in this case implies one equation~(\ref{FM-fin}).

Table~\ref{pardiso} shows comparison of the presented tensor product preconditioner approach and an alternative approach based on applying the Intel MKL Pardiso solver~\cite{MKL} for inverting the matrix $(H-E^*S)$ of~(\ref{FM-lin}) for IRAM iterations.
We note that although the matrix $(H-E^*S)$ is inevitably stored as a sparse matrix in CSR format in the Pardiso approach, the memory allocated for storing its elements values (the known CSR ``aa'' array of values) is freed right after the matrix has been factorized by Pardiso (phases 1 and 2 of Intel MKL Pardiso~\cite{MKL}).
In other words, we tried to do our best to minimize memory usage in the Pardiso approach.
However, as is seen from Table~\ref{pardiso}, memory and CPU time usage grow very fast as the number of basis functions increases.
\begin{table*}[!t]
\caption{\label{pardiso}Comparison of the tensor product preconditioner approach and the Pardiso-based approach. Two lower energy levels $2^1P$ and $3^1P$ (i.e. $J=1$, $\tau=1$ states with the parity with respect to exchange of electrons $p=1$) of the Helium atom have been requested in each run. The solution domains are (1: 36.0, 36.0; 3: 9.0, 6.0) a.u. The benchmark calculations~\cite{aznab15} energy values are -2.12384308649810135925 and -2.05514636209194353689 a.u.}
\centering
\begin{tabular}{p{2.5cm}|p{2.5cm}p{1.5cm}p{1.5cm}|p{1.5cm}p{1.7cm}p{1.5cm}|p{1.5cm}}
\hline
 & \multicolumn{3}{|l|}{Tensor product preconditioner} & 
\multicolumn{3}{|l|}{Pardiso} & \\
\cline{2-8}
Basis sizes & Total\newline memory/\newline $H-E^*S$/\newline Preconditioner\newline size, GB & Calculate\newline matrices/\newline IRAM\newline wall time, minutes & Total\newline CPU time,\newline minutes & Total\newline memory/\newline$H-E^*S$\newline size, GB & Calculate\newline matrices/\newline IRAM\newline wall time, minutes & Total\newline CPU time,\newline minutes & $E$, a.u. \\
\hline
(1: 22, 22, 9;\newline 3:19, 19, 12) & 0.14/0.09/0.01 & 0.01/0.3 & 2.1 & 2.29/0.28 & 0.05/0.6 & 2.7 & -2.12524\newline-2.05654 \\
(1: 31, 31, 9;\newline 3: 22, 22, 15) & 0.25/0.16/0.02 & 0.02/0.7 & 4.5 & 5.31/0.55 & 0.04/1.9 & 9.5 & -2.12365\newline-2.05495 \\
(1: 40, 40, 12;\newline 3: 25, 25, 18) & 0.45/0.30/0.04 & 0.05/1.2 & 7.3 & 12.5/1.07 & 0.1/6.9 & 35 & -2.12378\newline-2.05508 \\
(1: 49, 49, 12;\newline 3: 28, 28, 21) & 0.68/0.45/0.06 & 0.08/1.9 & 12 & 21.3/1.62 & 0.1/14 & 73 & -2.12385\newline-2.05513 \\
(1: 58, 58, 12;\newline 3: 28, 28, 21) & 0.85/0.56/0.10 & 0.1/2.3 & 15 & 29.4/2.04 & 0.2/22 & 120 & -2.12384\newline-2.05514
\end{tabular}
\end{table*}

\begin{figure}[!t]
\centering
\begin{minipage}[h]{0.49\linewidth}
\center{\includegraphics[width=0.99\textwidth]{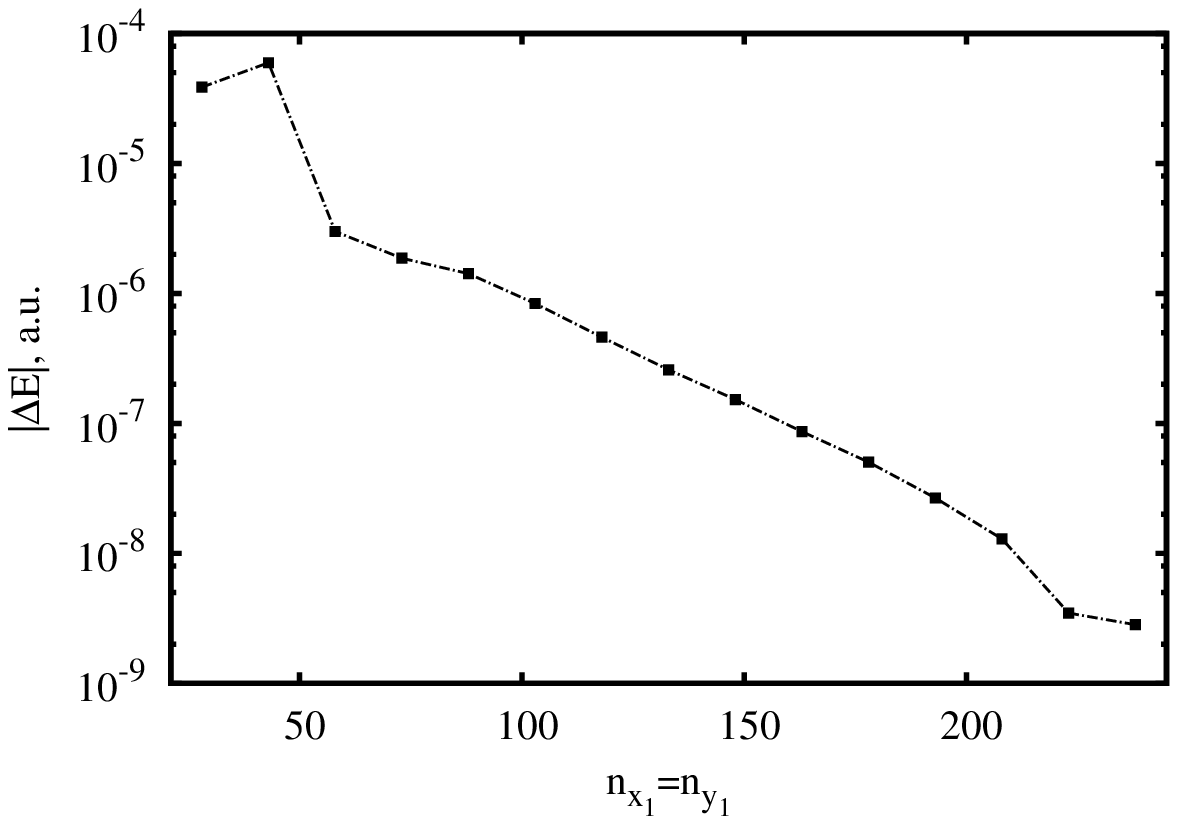}}\\
(a)
\end{minipage}
\hfill
\begin{minipage}[h]{0.49\linewidth}
\center{\includegraphics[width=0.99\textwidth]{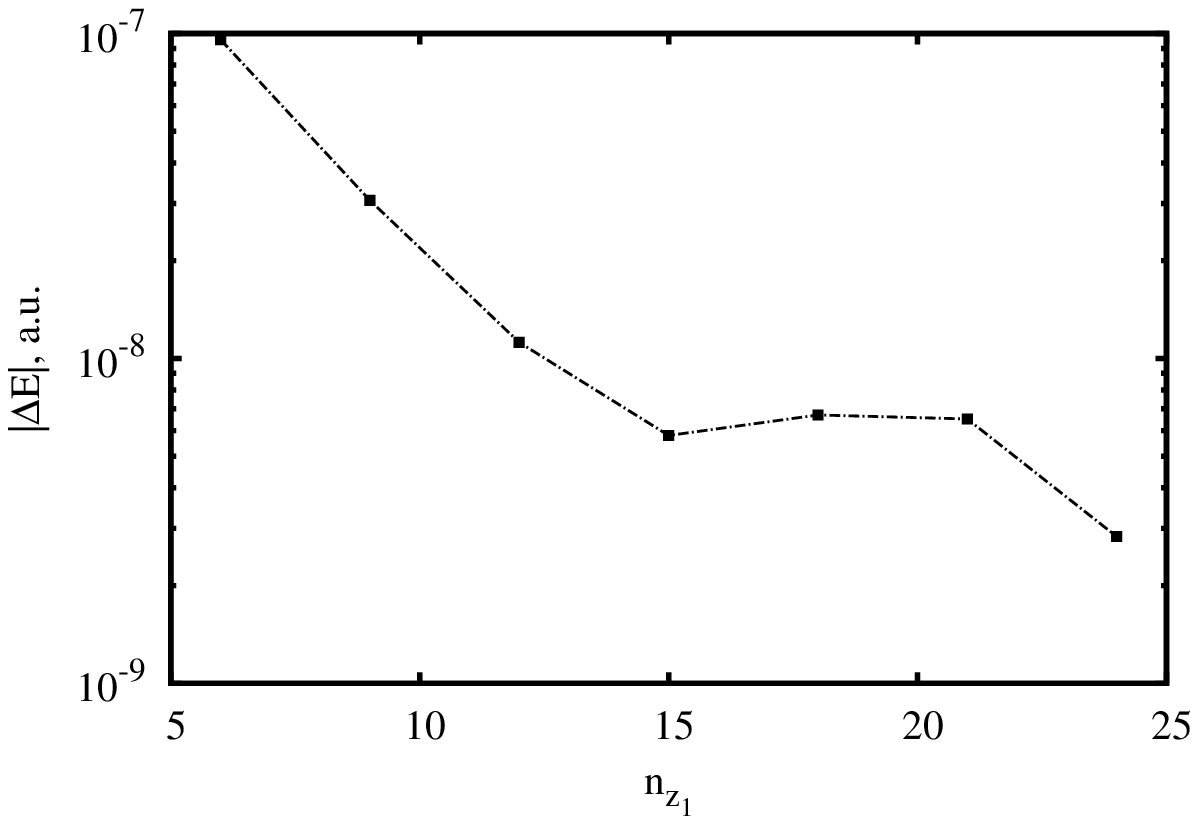}}\\
(b)
\end{minipage}
\hfill
\begin{minipage}[h]{0.49\linewidth}
\center{\includegraphics[width=0.99\textwidth]{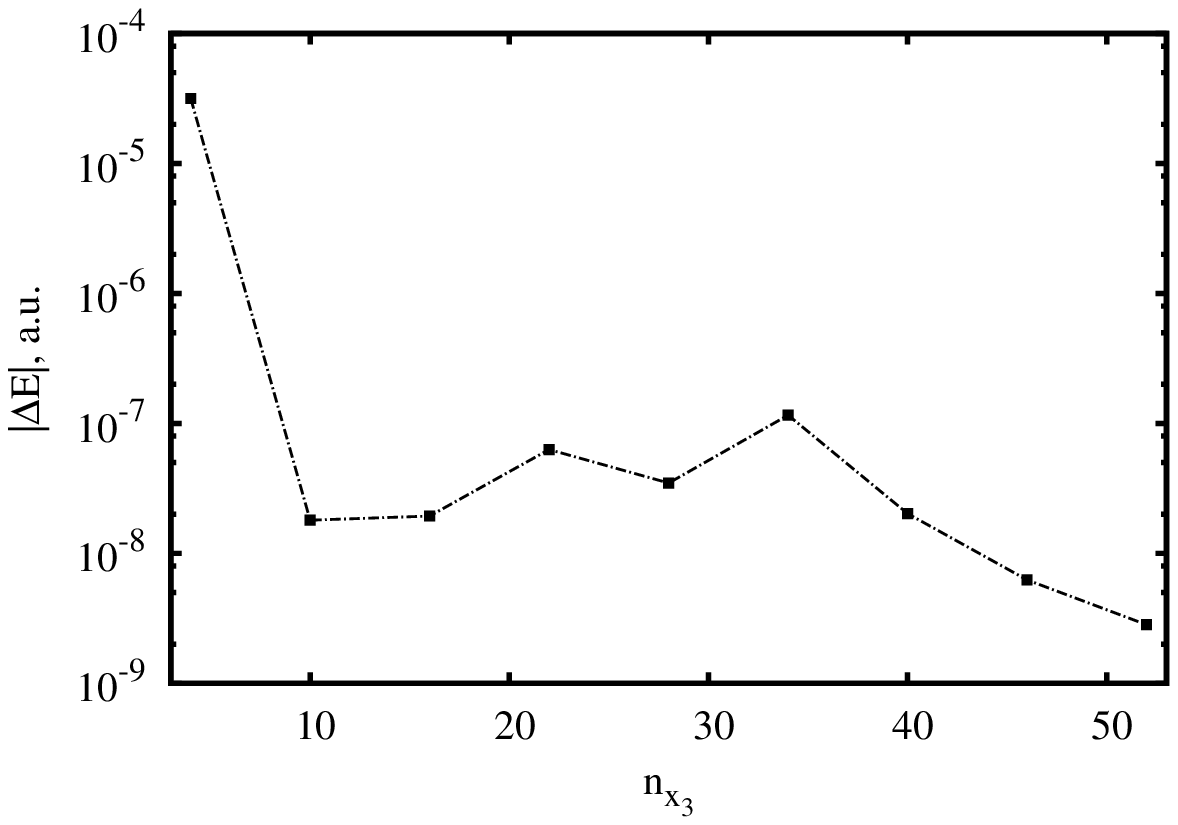}}\\
(c)
\end{minipage}
\hfill
\begin{minipage}[h]{0.49\linewidth}
\center{\includegraphics[width=0.99\textwidth]{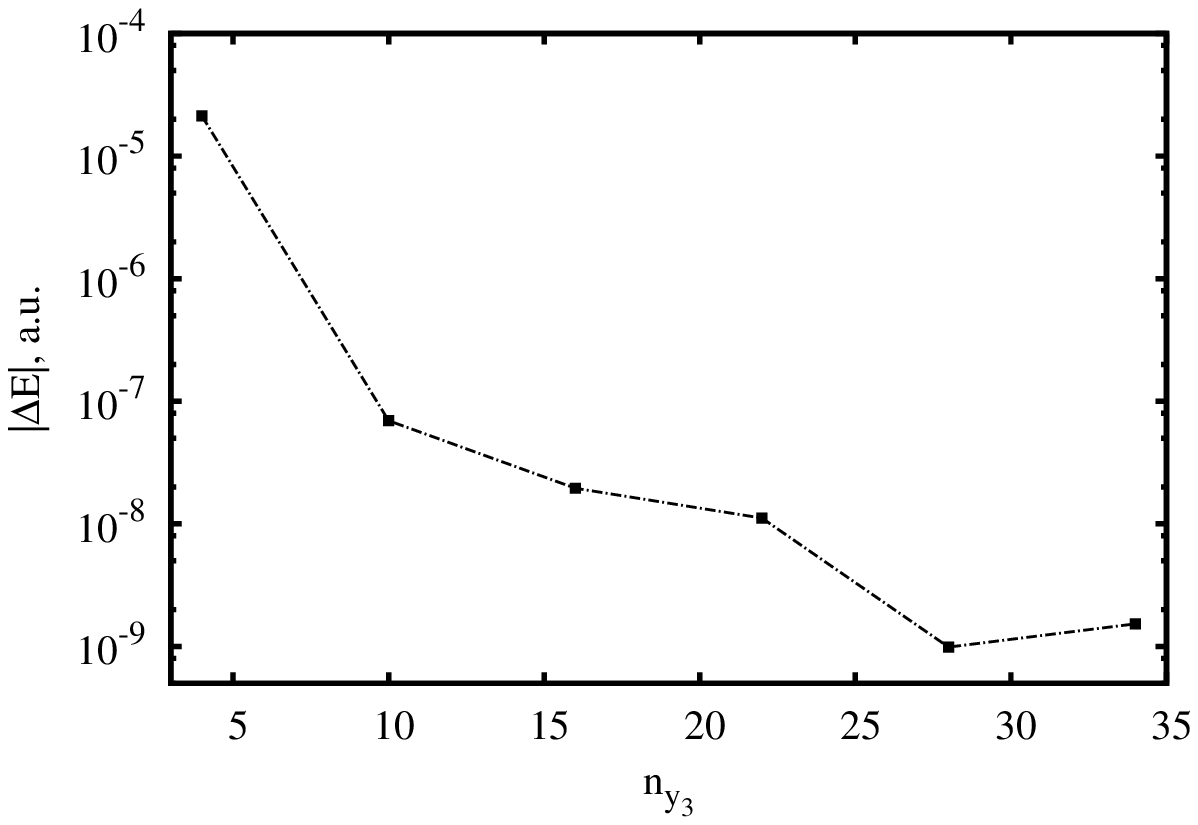}}\\
(d)
\end{minipage}
\hfill
\begin{minipage}[h]{0.49\linewidth}
\center{\includegraphics[width=0.99\textwidth]{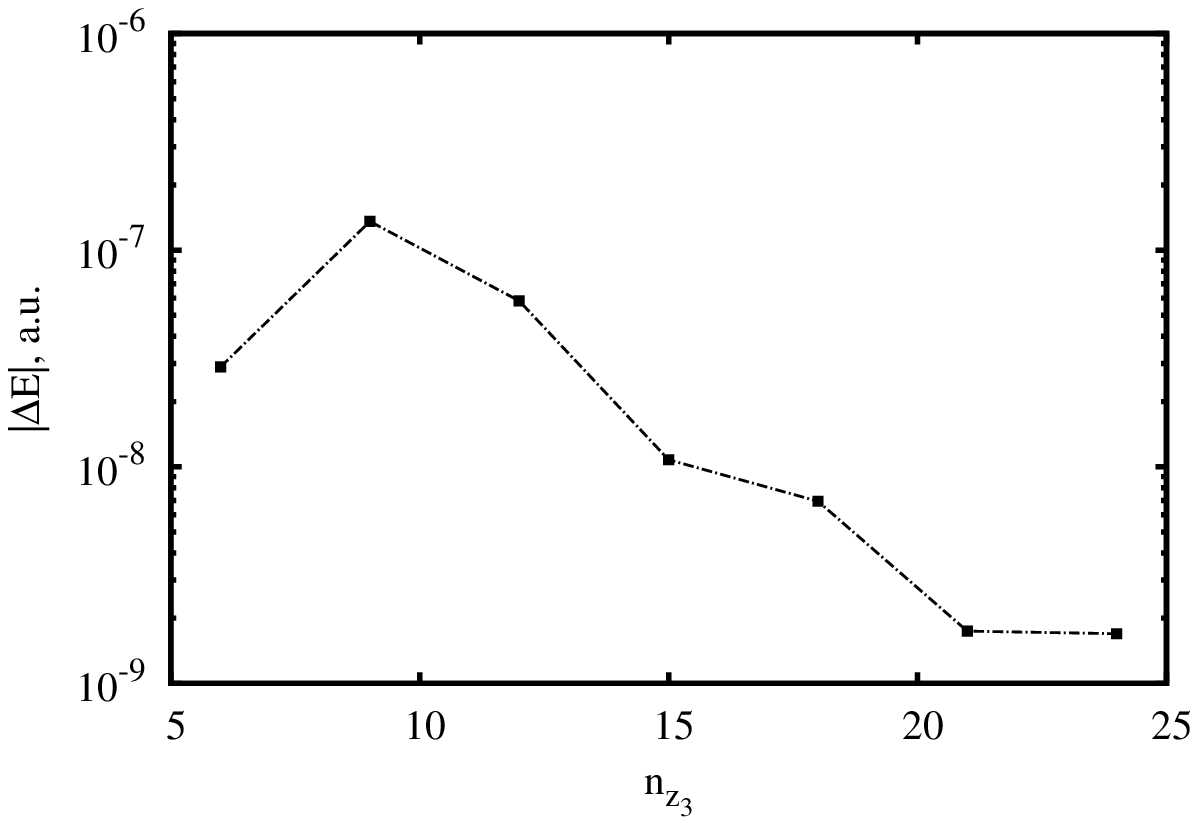}}\\
(e)
\end{minipage}
\caption{Convergence of $3^1D$ ($J=2$, $\tau=1$, $p=1$) energy level of atomic Helium with respect to basis sizes in different coordinates. In each figure the chosen parameter of the final basis (1: 238, 238, 24; 3: 52, 40, 27) is being changed. The solution domains are (1: 45.0, 45.0; 3: 11.0, 6.0) a.u. The difference $\Delta E$ is that of the calculated value and the benchmark value -2.05562073285224648939 of~\cite{aznab15}.
}
\label{He-conv}
\end{figure}

Fig.~\ref{He-conv} and Table~\ref{HeMol-conv} demonstrate convergence of energy level values of atomic Helium and molecular Helium trimer with respect to basis sizes. In the calculations, the largest presented basis sets lead to the matrices of linear size 4246848 for atomic Helium and 1342683 for molecular Helium.
In both cases, the agreement with benchmark values is good and our universal approach provides a clear convergence to these values. In the simpler case of atomic Helium the relative error of approximately $10^{-8}$--$10^{-9}$ is reached with basis size that is far from maximal possible one available for our calculations.

\begin{table*}[!t]
\caption{\label{HeMol-conv}Convergence of the ground state ($J=0$, $p=1$) energy level of Helium trimer with respect to basis sizes.
The solution domain is (1: 1200.0, 2450.0) a.u.
The benchmark calculations of~\cite{Roudn12} give value -4.00720e-7.}
\centering
\begin{tabular}{ll}
\hline
Basis sizes & $E$, $10^{-7}$ a.u. \\
(1: 58, 58, 12) & -2.9142\\
(1: 88, 88, 15) & -4.0895 \\
(1: 118, 118, 18) & -4.0219 \\
(1: 148, 148, 21) & -4.0070 \\
(1: 178, 178, 24) & -4.0081 \\
(1: 208, 208, 27) & -4.0081 \\
(1: 223, 223, 27) & -4.0077 \\
\hline

\end{tabular}
\end{table*}

Table~\ref{He-10} shows performance of the present paper algorithm for the high $J=10$ case. In this case a set of twenty coupled 3D PDEs~(\ref{FM-fin}) has been solved. Table~\ref{He-10} shows that the relative error of approximately $10^{-5}$ is reached with relatively small computer resources.
\begin{table*}[!t]
\caption{\label{He-10}Performance of the present paper algorithm for the high $J$ case. The ground and first excited state energies of the Helium atom with $J=10$, $\tau=1$, $p=1$ are calculated in each run. The solution domains are (1: 280.0, 280.0; 3: 2.0, 2.0) a.u.
}
\centering
\begin{tabular}{p{2.5cm}p{1.5cm}p{2.5cm}p{1.5cm}p{1.5cm}p{1.5cm}}
\hline
Basis sizes & Matrix\newline linear size & Total\newline memory/\newline $H-E^*S$/\newline Preconditioner\newline size, GB & Calculate\newline matrices/\newline IRAM\newline wall time, minutes & Total\newline CPU time,\newline minutes & $E$, a.u. \\
\hline
(1:58,58,6;\newline 3:4,4,6) & 223080 & 0.62/0.17/0.11 & 0.1/6.3 & 38 & -2.00398\newline-2.00332 \\
(1:88,88,9;\newline 3:7,7,9) & 771507 & 2.2/0.59/0.43 & 0.7/33 & 200 & -2.00412\newline-2.00346 \\
(1:118,118,9;\newline 3:7,7,9) & 1383327 & 4.1/1.06/1.01 & 1.9/60 & 350 & -2.00413\newline-2.00347 \\
(1:118,118,12;\newline 3:7,7,12) & 1844436 & 7.3/1.41/1.18 & 2.4/280 & 1500 & -2.00413\newline-2.00347 \\
\end{tabular}
\end{table*}

\section{Conclusions}
In this work we have presented the computational approach for solving the quantum three-body problem.
It is based on solving the Faddeev-Merkuriev equations in total orbital momentum representation.
These are the finite set of coupled three-dimensional PDEs, that  are solved by the method of spline collocations.
The idea of fast inverting the matrix of the operator at the left-hand side of the FM equations written in tensor product form is used.
In our approach it is done in the form of approximate preconditioning.
The tensor product form is also used in the matrix storage scheme, which leads to a significant economy in both computer resources and time under some assumptions on bases and collocation points.
In the calculations, we have shown the universality and efficiency of our approach in solving bound state problems of different nature.
In the future, we plan to use it for solving multichannel scattering problems in various systems of three neutral and charged particles.

\section*{Acknowledgments}
VAG' s work was supported by Russian Science Foundation grant No. 19-72-00076.
VAR, EAY and SLY' s work was supported by Russian Foundation for Basic Research grant No. 18-02-00492.
Research was carried out using computational resources provided by Resource Center "Computer Center of SPbU" (http://cc.spbu.ru).

\bibliographystyle{unsrt}
\bibliography{refs}

\appendix

\section{}
\label{app}
In this Appendix we clarify the assumptions under which the approximate equalities~(\ref{prop-eq}) hold.
Let we first introduce the eigenfunctions of the operator $\bm{d}_{z_\alpha}^{M'}$ defined in~(\ref{d-def})
\begin{equation}
\label{g-def}
g_{\ell}^{M'}(z_\alpha)=\sqrt{\frac{2\ell+1}{2}\frac{(\ell-M')!}{(\ell+M')!}}\frac{P_\ell^{M'}(z_\alpha)}{(1-z_\alpha^2)^{M'/2}},
\end{equation}
where $P_\ell^{M'}$ are the associated Legendre polynomials~\cite{NISTDLMF}.
From their properties it follows that~$g_\ell^{M'}\equiv0$ when $\ell<M'$, otherwise $g_\ell^{M'}$ is a polynomial of degree $\ell-M'$.
They satisfy
\begin{equation}
\label{g-ev}
\bm{d}_{z_\alpha}^{M'}g_{\ell}^{M'}(z_\alpha) = -\ell(\ell+1)g_{\ell}^{M'}(z_\alpha),
\end{equation}
\begin{equation}
\label{g-ud}
\bm{d}_{z_\alpha}^{M'\pm}g_\ell^{M'\pm1}(z_\alpha)=\mp\lambda^{\ell,\pm M'}g_\ell^{M'}(z_\alpha),
\end{equation}
with the operators $\bm{d}_{z_\alpha}^{M'\pm}$ from~(\ref{d-def}).

Now we are able to formulate the following statements:
\begin{enumerate}
\item If for a given $M'=M_0,\ldots,J$ the basis of functions $S_\alpha^k(z)$ is chosen so that it approximates the first $n_{z_\alpha}$ eigenfunctions $g_\ell^{M'}$ ($\ell=M',\ldots,M'+n_{z_\alpha}-1$) and their first two derivatives on the grid of collocation points $z_\alpha^\zeta$ chosen close to roots of $g_{M'+n_{z_\alpha}}^{M'}$, then the approximate equalities~(\ref{prop-eq}) hold.
\item If (nonlocal) basis functions $S_{\alpha}^k$ are polynomials of degree $k$ and roots of $g_{M'+n_{z_\alpha}}^{M'}$ are chosen as collocation points to represent equation~(\ref{FM-fin}) with given $M'$, then approximate equalities~(\ref{prop-eq}) are exact.
\end{enumerate}
To justify the statements, we note that since by definition~(\ref{z-dec}) the equality $D_{z_\alpha}^{M'}W_{z_\alpha}^{M'}=S_{z_\alpha}W_{z_\alpha}^{M'}\widetilde{\Lambda}_{z_\alpha}^{M'}$ holds, $\widetilde{\Lambda}_{z_\alpha}^{M'}$ is the matrix of approximate eigenvalues and columns of $W_{z_\alpha}^{M'}$ are the expansion coefficients in terms of basis functions $S_\alpha^k$ of approximate eigenvectors of the differential eigenvalue problem~(\ref{g-ev}), obtained via collocation procedure.
Then by assumption 1 the first approximate equality of~(\ref{prop-eq}) follows.
Now the approximate equality
\begin{equation}
\label{up}
D_{z_\alpha}^{M'-}W_{z_\alpha}^{M'-1} \approx S_{z_\alpha}W_{z_\alpha}^{M'\rightarrow}\Lambda_{z_\alpha}^{M'-\downarrow}
\end{equation}
is the consequence of exact equality~(\ref{g-ud}) with lower sign. Here the matrix
\begin{equation}
\Lambda_{z_\alpha}^{M'-\downarrow}=\mathrm{diag}\{\lambda^{M'-1,-M'},\lambda^{M',-M'},\ldots,\lambda^{M'+n_{z_\alpha}-2,-M'}\}
\end{equation}
and the matrix $W_{z_\alpha}^{M'\rightarrow}$ is obtained from $W_{z_\alpha}^{M'}$ by shifting its columns by one position to the right with replacing first column with zeros and throwing off the last column.
Similarly, the approximate equation
\begin{equation}
\label{down}
D_{z_\alpha}^{M'+}W_{z_\alpha}^{M'+1} \approx \big( S_{z_\alpha}W_{z_\alpha}^{M'\leftarrow}\big| \widetilde{g}_{M'+n_{z_\alpha}}^{M'} \big)\Lambda_{z_\alpha}^{M'+\uparrow}
\end{equation}
is the consequence of~(\ref{g-ud}) with upper sign.
The matrix
\begin{equation}
\Lambda_{z_\alpha}^{M'+\uparrow}=\mathrm{diag}\{-\lambda^{M'+1,M'},-\lambda^{M'+2,M'},\ldots,-\lambda^{M'+n_{z_\alpha},M'}\}
\end{equation}
and the matrix $W_{z_\alpha}^{M'\leftarrow}$ is obtained from $W_{z_\alpha}^{M'}$ by shifting its columns by one position to the left with throwing off the first one.
The column $\widetilde{g}_{M'+n_{z_\alpha}}^{M'}$ is filled with approximate values of the function $g_{M'+n_{z_\alpha}}^{M'}$ at collocation points.
By assumption the approximate equality $\widetilde{g}_{M'+n_{z_\alpha}}^{M'}\approx 0$ is true.
Then using~(\ref{up}),~(\ref{down}) and the definition~(\ref{z-dec}) of matrices $\overline{W}_{z_\alpha}^{M'}$, $W_{z_\alpha}^{M'}$
we obtain the remaining approximate equalities
\begin{eqnarray}
\overline{W}_{z_\alpha}^{M'}\left( D_{z_\alpha}^{M'-}W_{z_\alpha}^{M'-1} \right) & \approx &
\left( \overline{W}_{z_\alpha}^{M'} S_{z_\alpha} W_{z_\alpha}^{M'\rightarrow} \right) \Lambda_{z_\alpha}^{M'-\downarrow} =
\Lambda_{z_\alpha}^{M'-}, \nonumber \\
\overline{W}_{z_\alpha}^{M'}\left( D_{z_\alpha}^{M'+}W_{z_\alpha}^{M'+1} \right) & \approx &
\left( \overline{W}_{z_\alpha}^{M'} S_{z_\alpha} W_{z_\alpha}^{M'\leftarrow}\big|0 \right) \Lambda_{z_\alpha}^{M'+\uparrow} =
\Lambda_{z_\alpha}^{M'+}.
\end{eqnarray}
If a polynomial basis is used and collocation points are roots of polynomials $g_{M'+n_{z_\alpha}}^{M'}$, then $W_{z_\alpha}^{M'}$ is an exact representation of functions $g_\ell^{M'}$ in this basis and one has $\widetilde{g}_{M'+n_{z_\alpha}}^{M'}=0$.
Thus all approximate equalities above are exact.

\end{document}